\RequirePackage{fix-cm}
\documentclass[twocolumn,epjc3]{svjour3}  
\smartqed  % flush right qed marks, e.g. at end of proof
\RequirePackage{graphicx}
\journalname{Eur. Phys. J. C}

\usepackage{algorithm}
\usepackage{algpseudocode}\usepackage{listings}
\usepackage{xcolor}
\definecolor{backcolour}{rgb}{0.95,0.95,0.92}
\definecolor{codegray}{rgb}{0.5,0.5,0.5}
\lstdefinestyle{mystyle}{
    backgroundcolor=\color{backcolour},   
    numberstyle=\tiny\color{codegray},
    basicstyle=\ttfamily\footnotesize,
    breakatwhitespace=false,         
    breaklines=true,                 
    captionpos=b,                    
    keepspaces=true,                 
    numbers=none,                  
    numbersep=5pt,                  
    showspaces=false,                
    showstringspaces=false,
    showtabs=false,                  
    tabsize=2
}
\usepackage{fontawesome5}
\usepackage{hyperref}
\usepackage{biblatex}
\usepackage{caption}
\usepackage{subcaption}
\begin{document}
\sloppy

%\title{Fuzzing Frameworks for Server-side Web Security Testing: A Survey
\title{Fuzzing Frameworks for Server-side Web Applications: A Survey
%\thanksref{t1}
}
%\subtitle{Do you have a subtitle?\\ If so, write it here}

%\titlerunning{Short form of title}        % if too long for running head

\author{I Putu Arya Dharmaadi\thanksref{e1,addr1,addr2}
        \and
        Elias Athanasopoulos\thanksref{addr3}
        \and 
        Fatih Turkmen\thanksref{addr1} %etc.
}

%\thankstext{t1}{Grants or other notes
%about the article that should go on the front page should be
%placed here. General acknowledgments should be placed at the end of the article.
\thankstext{e1}{e-mail: arya.dharmaadi@rug.nl}

%\authorrunning{Short form of author list} % if too long for running head

\institute{University of Groningen, The Netherlands \label{addr1}
           \and
           Udayana University, Indonesia \label{addr2}
           \and
           University of Cyprus, Cyprus\label{addr3}
}

\date{Received: date / Accepted: date}
% The correct dates will be entered by the editor

\maketitle

\begin{abstract}
There are around 5.3 billion Internet users, amounting to 65.7\% of the global population, and web technology is the backbone of the services delivered via the Internet. To ensure web applications are free from security-related bugs, web developers test the server-side web applications before deploying them to production. The tests are commonly conducted through the interfaces (i.e., Web API) that the applications expose since they are the entry points to the application. Fuzzing is one of the most promising automated software testing techniques suitable for this task; however, the research on (server-side) web application fuzzing has been rather limited compared to binary fuzzing which is researched extensively. This study reviews the state-of-the-art fuzzing frameworks for testing web applications through web API, identifies open challenges, and gives potential future research. We collect papers from seven online repositories of peer-reviewed articles over the last ten years. Compared to other similar studies, our review focuses more deeply on revealing prior work strategies in generating valid HTTP requests, utilising feedback from the Web Under Tests (WUTs), and expanding input spaces. The findings of this survey indicate that several crucial challenges need to be solved, such as the ineffectiveness of web instrumentation and the complexity of handling microservice applications. Furthermore, some potential research directions are also provided, such as fuzzing for web client programming. Ultimately, this paper aims to give a good starting point for developing a better web fuzzing framework.
\end{abstract}

\section{Introduction} 
Fuzzing is an automated software testing technique that focuses on finding bugs, errors, or faults in the software under test (SUT) \cite{yun_fuzzing_2023} by creating many test cases in the form of malformed/semi-malformed inputs and feed them into the SUT without requiring human intervention. The inputs are produced by employing a variety of techniques (e.g., mutation) with the idea of triggering software vulnerabilities that manifest themselves in the form of a crash. Since fuzzing has excellent potential to discover security-related vulnerabilities \cite{zhu_fuzzing_2022}, fuzzing of binary applications has been studied extensively that led to the development of a plethora of binary fuzzers such as American Fuzzy Lop (AFL) \cite{Zalewski_American_2014} and libFuzzer \cite{libFuzzer_2016}, and a yearly competition~\cite{sbft2023}.

While the research on binary fuzzing has made impressive strides, the research on fuzz testing of web applications is only recently picking up and the existing work to date has been scattered. This is surprising because web applications are ubiquitous as most governments and companies provide their services through the World Wide Web that allows access from a plethora of devices including desktop computers, mobile phones and tablets. According to a study by Cloudflare in December 2021~\cite{cloudflare2021}, around 25\% of all API traffic in its network is web related traffic and is twice as likely to be blocked than other API requests. More strikingly, the development and maintenance of web applications has a huge market size (56B\$ in 2021 and is expected to rise up to 89B\$ by 2027 \footnote{https://www.businessresearchinsights.com/market-reports/web-development-market-109039}) so the testing of web applications for security vulnerabilities has a significant commercial value.

Web application fuzzing has its own challenges, and many techniques from binary fuzzing are not directly applicable. For instance, web APIs only accept valid HTTP requests to be executed, which means that the malformed test cases commonly produced by fuzzers will be rejected by web servers. This condition made some of the existing web API fuzzing frameworks utilise OpenAPI specifications to create templates for HTTP requests which are then rendered with the correct values to form valid HTTP request sequences. However, there is a lack of research exploring other techniques to reach all hidden statements in web applications. Another challenge is that no proper benchmark for the experimental evaluation of web application fuzzers exists. 

In this paper, we conduct a review to summarise the existing works on web application fuzzing while analysing their strengths and limitations. From a more practical point of view, our study reveals how prior works designed strategies to generate HTTP requests, utilise feedback from the web under test (WUT), and expand relevant input spaces, which are all crucial for the effectiveness of fuzzing. Finally, we identify five areas of challenges that need to be addressed in order to improve fuzzing effectiveness and efficiency in web applications and four potential research directions that can be massively studied in the future.

\subsection{Research Questions}
In organizing our paper, we employed the following research questions (RQ), which are grouped according to various objectives.
    \begin{itemize}
        \item Producing valid HTTP requests
        \begin{itemize}
            \item \textbf{RQ1}: What techniques are used to generate HTTP request templates?
            \item \textbf{RQ2}: How are the request templates rendered?
        \end{itemize}
        \item Feedback observation
        \begin{itemize}
            \item \textbf{RQ3}: What kind of feedback is extracted from the WUT?
            % \item \textbf{RQ4}: What kind of instrumentation methods are used for the WUT?
            \item \textbf{RQ4}: How are the WUTs instrumented?
            \item \textbf{RQ5}: What vulnerabilities are observed?
        \end{itemize}
        \item Input space expansion
        \begin{itemize}
            % \item \textbf{RQ6}: What techniques are used to expand the existing input space?
            \item \textbf{RQ6}: How are the existing input spaces expanded?
        \end{itemize}
        \item Empirical evaluation
        \begin{itemize}
            \item \textbf{RQ7}: What benchmarks are used for empirical evaluations?
        \end{itemize}
        \item Open challenges
        \begin{itemize}
            \item \textbf{RQ8}: What open challenges are identified?
        \end{itemize}
    \end{itemize}

\subsection{Scope, Related Works, and Contributions}
\label{related-work}
Web application fuzzing is similar to web vulnerability scanning in terms of their approaches. Since we aim to review fuzzing strategies, this section first clarifies the differences between the fuzzer and the scanner, in general, to make it clear. In addition, since the fuzzing-related review topic is packed, this section then stresses the uniqueness of this work compared to other survey papers. Finally, a summary of our contributions is provided in the last part.

\subsubsection{Fuzzer vs vulnerability scanner}
Both fuzzers and vulnerability scanners work automatically to find software vulnerabilities. However, there are certain differences between as we discuss in this section. While a fuzzer produces plenty of malformed/semi-malformed inputs to make a software crash and let the software developers identify any vulnerability (mostly \textbf{0-day vulnerability}) behind the crash \cite{liang_fuzzing_2018} \cite{klooster_continuous_2023}, a scanner scans and injects available APIs with malicious payload for finding \textbf{pre-defined vulnerabilities}, such as SQL injection and XSS \cite{amankwah_empirical_2020}. Nevertheless, this distinction becomes increasingly blurred because, recently, there have also been vulnerability-driven fuzzers aiming to find certain vulnerabilities rather than just a crash (explained in Section \ref{vulnerability-driven-fuzzing}). Another difference is that a fuzzer performs dynamic testing using either a black-box, grey-box, or white-box approach \cite{chen_systematic_2018}; meanwhile, the dynamic scanner works from outside of the target (i.e., only uses the black-box approach) as per OWASP description~\cite{noauthor_vulnerability_2024}. In a nutshell, the fuzzer may have a broader usage than the scanner. This survey focuses more on fuzzing, which can be black-box, grey-box, or white-box, used for web application testing.

\subsubsection{Related works}
%Several works have been identified that are similar to this study in the last ten years. 
We identified several works that are similar to this study in the last ten years. Based on their focus, those works can be grouped into either general fuzzing or web API testing. First, some survey papers focused on the fuzzing approach in general, which is not explicitly intended for a specific platform, so their focus is different from ours. For example, the work of Chen et al. \cite{chen_systematic_2018} and Li et al. \cite{li_fuzzing_2018} focused on exploring techniques for improving fuzzing in general. Another example is the work of Zhu et al. \cite{zhu_fuzzing_2022}, which reviewed the knowledge gaps of general fuzzing. Instead of reviewing fuzzing in general, our work explores \textbf{the fuzzing approaches} (i.e., grammar-based and mutation-based) \textbf{tailored for web applications}, which utilises web resources that cannot be found in other application domains. Second, several works are done to review fuzzers for web API, such as the work of Martin-Lopez et al. \cite{martin-lopez_online_2022}, Zhang et al. \cite{zhang_open_2023}, and Golmohammadi et al. \cite{golmohammadi_testing_2023}. Their works are quite similar to ours; however, they reviewed using a general web API testing perspective, which does not explicitly focus on fuzzing strategies (see Table \ref{tab:survey-paper-comparison}). Our survey focuses on revealing prior work strategies seen from the fuzzing approaches: \textbf{generating valid HTTP requests, utilising feedback from the WUTs, and expanding relevant input spaces}.

\subsubsection{Contribution}
To conclude, our study's contributions are as follows.
\begin{enumerate}
    \item We review existing fuzzing studies specially designed for \textbf{web application testing through web API}. We investigate their strategies for \textbf{input generation, mutation, and feedback utilisation}.
    \item We analyse benchmarks used mostly for empirical evaluations.
    \item We identify the remaining challenges that need to be solved.
    \item We give some insights for future research directions.
\end{enumerate}

\begin{table*}
    \caption{Comparison of this survey paper to other similar reviews in web API testing.}
    \label{tab:survey-paper-comparison}
    \renewcommand{\arraystretch}{1.5}
%    \begin{tabular*}{\textwidth}{@{\extracolsep{\fill}}llp{4cm}p{8cm}@{}}
    \begin{tabular*}{\textwidth}{@{\extracolsep{\fill}}p{1cm}p{4cm}p{11cm}@{}}
        \hline
%         Paper & Group & Main Focus & Samples of the Results\\
         Paper & Main Focus & Samples of the Results\\
         \hline
         \cite{martin-lopez_online_2022} %& Web API testing 
         & Black-box API testing (e.g., failure detection capability) & Experimental results on failure detection and fault detection capabilities in 13 online APIs under test\\
         \cite{golmohammadi_testing_2023} %& Web API testing 
         & REST API testing approaches in general & Metrics for evaluating API testing effectiveness (e.g., coverage), testing techniques (e.g., black-box and white-box), kind of testing (e.g., system testing)\\
         \cite{zhang_open_2023} %& Web API testing 
         & Empirical assessment and open technical problem analysis & Empirical comparison and technical analysis of seven state-of-the-art web API fuzzers \\
         This paper %& Web API fuzzing 
         & Adjusted fuzzing strategies for Web API testing & Diverse techniques for generating valid HTTP requests (Section \ref{request-template-generation} and \ref{template-rendering}), utilising feedback from the WUTs (Section \ref{execution-and-feedback}), and expanding relevant input spaces (Section \ref{mutation-strategies})\\
         \hline
    \end{tabular*}
\end{table*}

\section{Background}
This section explains server-side web applications and the basic theory of fuzzing.

\subsection{Server-side Web Application and Web API}
\label{server-side-web}
\begin{figure}
    \centering
    \includegraphics[width=1 \linewidth]{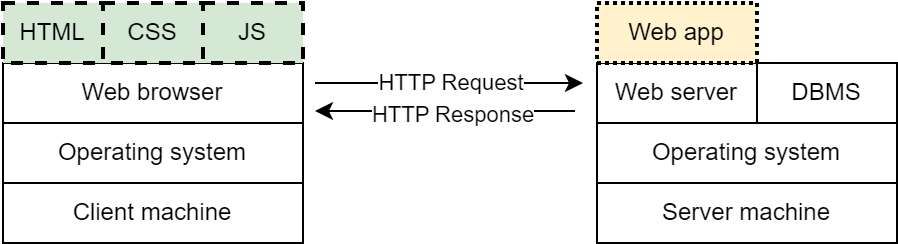}
    \caption{Web application overview. The dashed boxes are called client-side web applications, while the dotted ones are called server-side web applications or only "web applications" for ease of reference.}
    \label{fig:web-architecture}
\end{figure}

The server-side web application is the application running on the web server that executes any requests sent by the client \cite{li_survey_2014} (see the illustration in the Figure \ref{fig:web-architecture}). It works together with the web server to filter out broken or malicious request formats. We then only use the term "\textbf{web application}" to refer to server-side web applications for ease of reference. 

The client sends requests to the web application through the \textbf{web API}, an interface enabling other users, like humans or programs, to access the web application functions through the computer network \cite{jin_web_2018}. The most popular paradigm for accessing HTTP-based web API is Representational State Transfer (RESTful) \cite{presutti_restful_2014}. RESTful is an architectural style in the web application to represent a standard interface that enables the client to interact with or manipulate specific resources \cite{fielding2000architectural}. RESTful APIs are supposed to be stateless; however, since they are connected to stored systems, such as databases or cache, they can be seen as stateful systems. 

As explained below, researchers utilised several web API attributes to develop a web API fuzzing framework.

\subsubsection{HTTP Methods and Response Codes}
\label{http-method-response-code}
Clients request the web API using various HTTP methods defined in the HTTP standard. Consequently, fuzzing researchers considered this rule to develop a test case generator producing valid input data. The four most commonly used methods related to web resource management are \texttt{GET}, \texttt{POST}, \texttt{PUT}, and \texttt{DELETE} \cite{richardson_restful_2013}. After sending the request using one of the methods, the web server replies to the client with an HTTP response code. It is a three-digit code produced by the web application to help a client generally know what happened after processing the client request \cite{richardson_restful_2013}. Most web fuzzers utilise this information to decide whether their mutated input triggered a particular web behaviour. The most common response codes are explained as follows.
\begin{itemize}
    \item 2xx (OK): describing successful processing of the request and the web replies with intended resources.
    \item 4xx (Bad Request): describing the client's request may be incorrect or malformed.
    \item 5xx (Internal Server Error): describing a problem happening in the web application caused by the client's request.
\end{itemize} 

\subsubsection{Open-API Specification}
\label{open-api-specification}
To help software developers comprehend web API usage, web API developers should provide standard documentation. Many tools and technologies related to API documentation are available, making the developer easy in preparing the documentation, such as RESTful API Modeling Language (RAML)\footnote{http://raml.org/}, Swagger \footnote{http://swagger.io/}, API Blueprint\footnote{https://apiblueprint.org/}, and others \cite{de_api_2017}. OpenAPI Specification, formerly known as Swagger, has been the official standard for documenting web RESTful API because of its mass adoption \cite{noauthor_get_nodate}. Therefore, most fuzzing frameworks focused on applications with an OpenAPI specification to generate valid test cases. The example of the OpenAPI specification with some essential components in version 3.1.0 written in YAML format is listed in \ref{appendix:openapi-example}.

The crucial points most concerned by the web API fuzzer are the endpoint, method, and parameter. Endpoint (line 5 of \ref{appendix:openapi-example}) is a path string pointing to a specific URL of the API that can be followed by a token and/or an argument \cite{lin_forest_2023}. Method (line 6) refers to the HTTP methods like GET or POST, and the Parameter (lines 7-15) is data formed key-value pairs sent to the API server. One endpoint can be called using different methods, so the API server treats each endpoint-method combination differently.

\subsection{Fuzzing}
\label{fuzzing-definition}
Fuzzing, standing for fuzz testing, is an automatic software testing intended to find vulnerabilities or bugs, first proposed by Miller \emph{et al.} \cite{miller1990empirical} in 1988. The main ideas of this testing are automatically producing a huge amount of input data, injecting them into the software under test (SUT), and then watching the software's behaviour, whether it results in a crash, fault, or hang \cite{liang_fuzzing_2018}. Algorithm \ref{fuzzing-algorithm} illustrates the fuzzer process using the \textbf{mutation-based approach} (explained in Section \ref{mutation-input}). First, a fuzzer calls a mutation method to pick a random seed from the corpus and to produce an input (line 5). Then, the fuzzer injects the input into SUT and gets feedback (line 6). Lastly, the fuzzer stores the input if the feedback contains an error or is interesting (lines 7-13). The original fuzzing approach proposed in 1988 has become extremely competent with recent developments. Instead of producing genuinely random inputs, additional techniques were proposed to make it more intelligent in finding error-triggering inputs, such as program instrumentation \cite{li_fuzzing_2018}. Based on the available knowledge about the SUT (e.g., source code), fuzzing can be classified into three categories: Black-box, Grey-box, and White-box \cite{chen_systematic_2018}. While the fuzzing model running without having access to the SUT's source code was grouped into the black-box fuzzing, the rest that have varying levels of information are classified as either grey-box or white-box fuzzing. If the fuzzing methods only instrument the code for obtaining coverage information at runtime, they are considered the grey box. Otherwise, they are white-box.

\begin{algorithm}
\caption{Mutation-based fuzzer process, extracted from \cite{andreas_zeller_fuzzing_2023}}\label{fuzzing-algorithm}
\begin{algorithmic}[1]

\Require $s, I$   \Comment{Set the software under test and samples of the input}
%\Output $B, J$

\State $O \gets \emptyset$  \Comment{Initialize Output set}
\State $B \gets \emptyset$  \Comment{Initialize Bug set}
%\State $COV \gets \emptyset$  \Comment{Initialize Coverage set}
\State $C \gets I$  \Comment{Corpus set is filled with input sample}

\While{$true$}
\State $i \gets Mutate (C)$     \Comment{Get an input from mutation process}
\State $feedback \gets Inject(i,s)$   \Comment{Get the feedback from the execution}

\If{$feedback$ contains error}
    \State $C \gets C \cup i$      \Comment{Save the input into Corpus}
    \State $O \gets O \cup i$      \Comment{Save the input into Output}
    \State $B \gets B \cup o$      \Comment{Save the output into Bug}
\ElsIf{$feedback$ is interesting}   \Comment{e.g., feedback score is high}
    \State $C \gets C \cup i$      \Comment{Save the input into Corpus}
%    \State $COV \gets COV \cup path$      \Comment{Save the path into Coverage}
\EndIf
\EndWhile
\\ \Return $B, O$
\end{algorithmic}
\end{algorithm}

The following sections will explain the common fuzzing approaches used in web API testing.

\subsubsection{Mutation-based Input Generation}
\label{mutation-input}
Mutation-based fuzzing is the most common fuzzing technique that takes valid input data and then creates new inputs by making small, random changes (mutations) to the input \cite{manes_art_2021}. Several common mutations can be applied to the input data, such as flipping some bits, inserting or deleting characters, and changing the order of existing bytes. This mutation-based input is straightforward because it does not require any knowledge of the software under test and can be used in a black-box testing scenario. However, since the quality of the initial inputs significantly influences the fuzzing performance, having a diverse set of well-formed seed inputs is crucial to begin the fuzzing process \cite{godefroid_fuzzing_2020}. That kind of input can help fuzzer to explore more execution paths quickly. In the context of web API testing, fuzzing is adjusted to adopt more mutation strategies. The mutation can change the HTTP request order in a sequence, HTTP request structure, or HTTP parameter values. These techniques will be explained in more detail in Section \ref{mutation-strategies}. 

\subsubsection{Grammar-based Input Generation}
\label{grammar-based-input}
Mutation-based fuzzing demands at least one initial valid input from the user and mutates it to produce plenty of input. However, recent development shows that a fuzzer can employ another way to work as the input generator. It is grammar-based fuzzing, an approach to generate valid input data by employing a grammar that specifies the structure and constraints of the input data \cite{manes_art_2021}. Other researchers, Pezze and Young (\cite{pezzè2008software}, called this approach specification-based testing because it generates test cases based on the test specification (i.e., grammar). In certain applications, grammar creation is greatly assisted by supporting documents that are an integral part of the domains. For example, in the web API context, to produce valid API requests, a fuzzer can employ a grammar model deduced from the OpenAPI Specification or HTML documents (explained in Section \ref{request-template-generation}). The grammar created in this context consists of request type, server endpoint, and required parameter names, which are set up as static, and required parameter values that are \textit{fuzzable} or replaceable (see \ref{appendix:grammar-example-by-restler}). So, when calling this grammar to build one valid HTTP request, the fuzzer copies the static data and changes the fuzzable data to concrete values that can be obtained from various sources, such as a dictionary \cite{viglianisi_resttestgen_2020}. Implementing grammar-based input generation helps the fuzzing to produce HTTP requests that strictly match the rules or specifications.

\section{Survey Methodology}
\label{survey-methodology}
This section presents the methodology used in collecting and filtering the paper in the literature search.

\subsection{Searching Papers}
\label{paper-searching}
We review papers issued in seven research article repositories: ACM Digital Library, IEEE Explore, Science Direct, Wiley, Web of Science, MIT Libraries, and Springer because those are well-known publication venues storing top articles on computer science topics. 
In addition, most similar survey papers we mentioned in Section \ref{related-work} used those for the article sources. Since our scope is existing fuzzers specially designed for web application testing through web API, we used keywords, namely \texttt{fuzzing}, \texttt{web}, \texttt{REST}, \texttt{RESTful}, \texttt{API}, and \texttt{testing}, and formulated them into different query syntax (see Table~\ref{tab:paper-collection}) to search relevant papers specifically on the repositories. We limited our search to the last ten years (2013--2023) to obtain the most recent studies. We applied additional filters in the repositories to show only research articles. We also paid attention to double-indexed articles (e.g., indexed in ACM and IEEE) to count them as one.

\begin{table*}
    \caption{Number of papers collected from some repositories after manual reduction (sections \ref{manual-paper-selection}).}
    \centering
    \begin{tabular*}{\textwidth}{@{\extracolsep{\fill}}lp{2cm}p{6cm}p{4cm}p{1cm}@{}}
        \hline
         No & Repository & Query & Additional filter & Selected papers\\
         \hline
         1 & ACM Digital Library & [All: fuzzing] AND [All: web] AND [All: api] AND [All: testing] AND [All: rest or restful] AND [E-Publication Date: (01/01/2013 TO *)] & Only "Research Article" & 14\\
         2 & IEEE Explore & "Full Text \& Metadata":fuzzing and web and api and testing and rest or restful & Only "Conferences" and "Journals" \& Publication time: 2013-2023 & 18\\
         3 & Science Direct & Terms: fuzzing and web and api and testing and rest or restful & Only "Research Articles" \& Publication time: 2013-2023 & 2\\
         4 & Wiley & Anywhere: fuzzing and web and api and testing and rest or restful & Only "journals" \& Publication time: 2013-2023 & 1\\
         5 & Web of Science & ((((ALL=(fuzzing)) AND ALL=(web)) AND ALL=(api)) AND ALL=(testing)) AND ALL=(rest or restful) & Only "journals" \& Publication time: 2013-2023 & 1\\
         6 & MIT Libraries & fuzzing and web and api and testing and rest or restful & Only "journal articles" \& Publication time: 2013-2023 & 0\\
         7 & Springer & fuzzing and web and api and testing and rest or restful & Only "conference paper" or "article" \& Publication time: 2013-2023 & 2\\
         \hline
         \multicolumn{4}{c}{Total} & 38 \\
         \hline
    \end{tabular*}
    \label{tab:paper-collection}
\end{table*}

\subsection{Manual Filtering of the Collected Papers}
\label{manual-paper-selection}
Papers whose main ideas do not propose or improve a fuzzer must be dropped. To remove such papers from the paper collection, we scanned the paper abstract, experiment, and result sections. We paid attention to those sections because papers focusing on developing web API fuzzers generally tested their frameworks for any web application and showed their experimental results. If there is no experimental result, we consider the paper not to propose a new improvement method and then drop it from the collection. Specifically, papers were excluded because of one of the following reasons:
\begin{enumerate}
    \item They were survey or review papers.
    \item They only conducted empirical studies to compare other studies without proposing a new framework.
    \item Instead of a web application, they intended to test the browser, applications, or other engines.
    \item They intended to test specific application APIs (e.g., machine learning APIs or interpreter APIs).
    \item They do not do server-side web application testing but rather client-side testing or UI testing.
    \item They mentioned web API testing and fuzzing, but their main topic is not web API testing techniques (e.g., they conduct short-campaign testing for analysis purposes).
\end{enumerate}

\subsection{Collection Expansion}
After performing the methodology explained in prior sections, we ended up with 38 articles (see Table \ref{tab:paper-collection}). To reduce the chance of missing relevant works, we further scanned the related work sections of these 38 papers. In addition to broadening our search, we checked other studies that had cited the collected papers and reviewed them to determine whether they were related to web API fuzzing. If affirmative, we include them in our paper collection. 

\subsection{Summary on Publications}
\label{summary-publication}
\begin{figure}
    \includegraphics[width=1\linewidth]{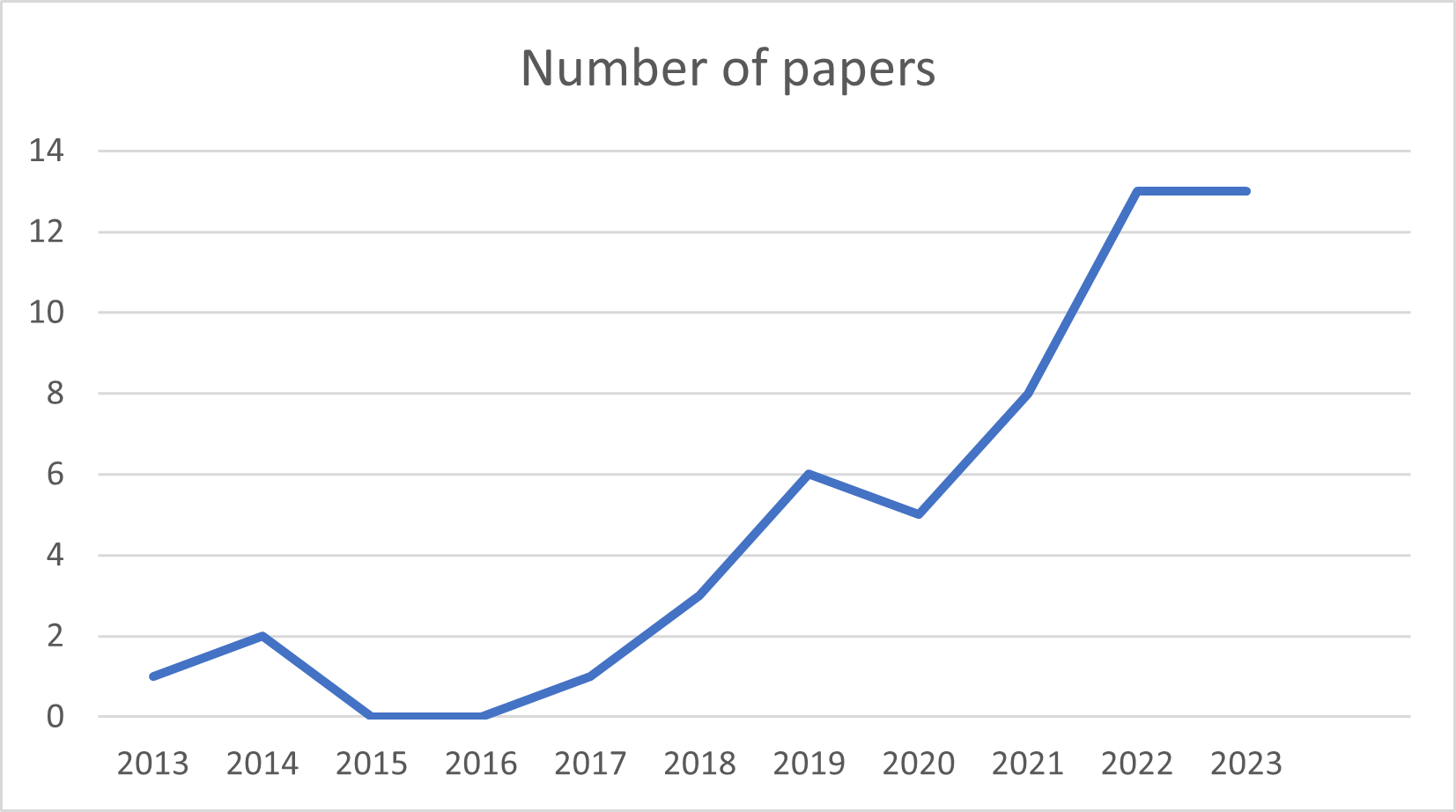}
    \caption{Number of papers in our collection published over the last ten years.}
    \label{fig:number-of-paper}
\end{figure}

Finally, we obtained 53 articles, including 38 initially found by search and 15 new ones by performing the collection expansion step. The majority of them (64\%) are conference papers published in proceedings, while the rest are journal articles (32\%) and book chapters (4\%). Designing a web API fuzzing framework is a trending topic because more and more papers are published over time (see Figure \ref{fig:number-of-paper}). Evidently, there were only 3 papers in total in the initial four-year period, but after that, the number of publications increased significantly.

\section{Web API Fuzzing Overview}
Generally, all reviewed studies defined their works as (web) API fuzzing, in which these two phrases---API and fuzzing---have been explained in Sections \ref{server-side-web} and \ref{fuzzing-definition}, respectively. Based on the descriptions in those sections, we conclude that web API fuzzing is an automatic test through typical web interfaces with a huge amount of HTTP requests for finding web vulnerabilities. Even though it is related to the network protocol, web API fuzzing completely differs from network fuzzing because the targeted applications differ. For the same reason, web API fuzzing differs from general/binary fuzzing. Table \ref{tab:different-fuzzing-comparison} explains the difference between web API fuzzing and other similar fuzzing types.

Based on the approach to finding vulnerabilities, existing works on web API fuzzing are classified into standard and vulnerability-driven fuzzing. While the former is intended to test all WUT codes without considering certain vulnerabilities to look for, the latter is supposed to test specific code regions with pre-defined rules to examine specific vulnerabilities the user is looking for. 
Those groups are explained more in the sections below.

\begin{table}
    \caption{Comparison between web API fuzzing and other similar kinds of fuzzing.}
    \centering
%    \begin{tabular*}{\textwidth}{@{\extracolsep{\fill}}lll@{}}
    \begin{tabular}{@{\extracolsep{\fill}}llp{5cm}@{}}
        \hline
         No & Name & Software Under Test \\
         \hline
         1 & (Binary) Fuzzing & Binary (compiled) applications for Desktop machines \\
         2 & Network Fuzzing & Server applications (e.g., LightFTP) \\
         3 & Web API Fuzzing & Server-side web applications tested through web API \\
         \hline
    \end{tabular}
    \label{tab:different-fuzzing-comparison}
\end{table}

\subsection{Standard Fuzzing}
\label{standard-fuzzing}
This group is called standard fuzzing because it works like the initial fuzz testing proposed by Miller \emph{et al.} \cite{miller1990empirical}. The standard fuzzer focuses on producing plenty of test cases leading to the WUT crashes, and then, based on the logs of raised errors, the tester analyses the vulnerabilities, errors, or bugs behind the crash. Since it is not designed to look for specific bugs or vulnerabilities in the WUTs, the fuzzer mutates the input with standard mutation rules. Its main objective is to reach and test all statements and possible states in the web application. For those reasons, the standard fuzzing has a high chance of finding the 0-day vulnerabilities. Some examples of existing works that purely design the standard web API fuzzing are Restler \cite{atlidakis_restler_2019} and MINER \cite{lyu_miner_nodate}. On the other hand, there are previous studies that initially built the standard fuzzing yet combined with specific rules to trigger specific vulnerabilities. They are still classified in the standard fuzzing because they focus on finding crashes and testing all lines of code; triggering the specific vulnerabilities can be an additional feature. 

\subsection{Vulnerability-driven Fuzzing}
\label{vulnerability-driven-fuzzing}
Unlike the first group, which focused on testing the entire code, this group may only look for certain code regions containing desired vulnerabilities. Because its work is designed to focus on specific vulnerabilities, this group is called vulnerability-driven fuzzing. The web API fuzzing frameworks that belong to the vulnerability-driven fuzzing are already supplied with custom mutators and vulnerability checkers to achieve their targets. The fuzzing framework uses particular mutators to produce input satisfying the requirement for triggering the vulnerabilities in the code. Then, it checks for the symptoms of vulnerability using the custom checkers. Some examples of existing web API fuzzers working in this category are Cefuzz \cite{zhao_cefuzz_2022} and Zokfuzz \cite{zhang_zokfuzz_2022}.

\section{Web API Fuzzer Workflow}
\label{review-result-restful-api}
\begin{figure}
    \includegraphics[width=1\linewidth]{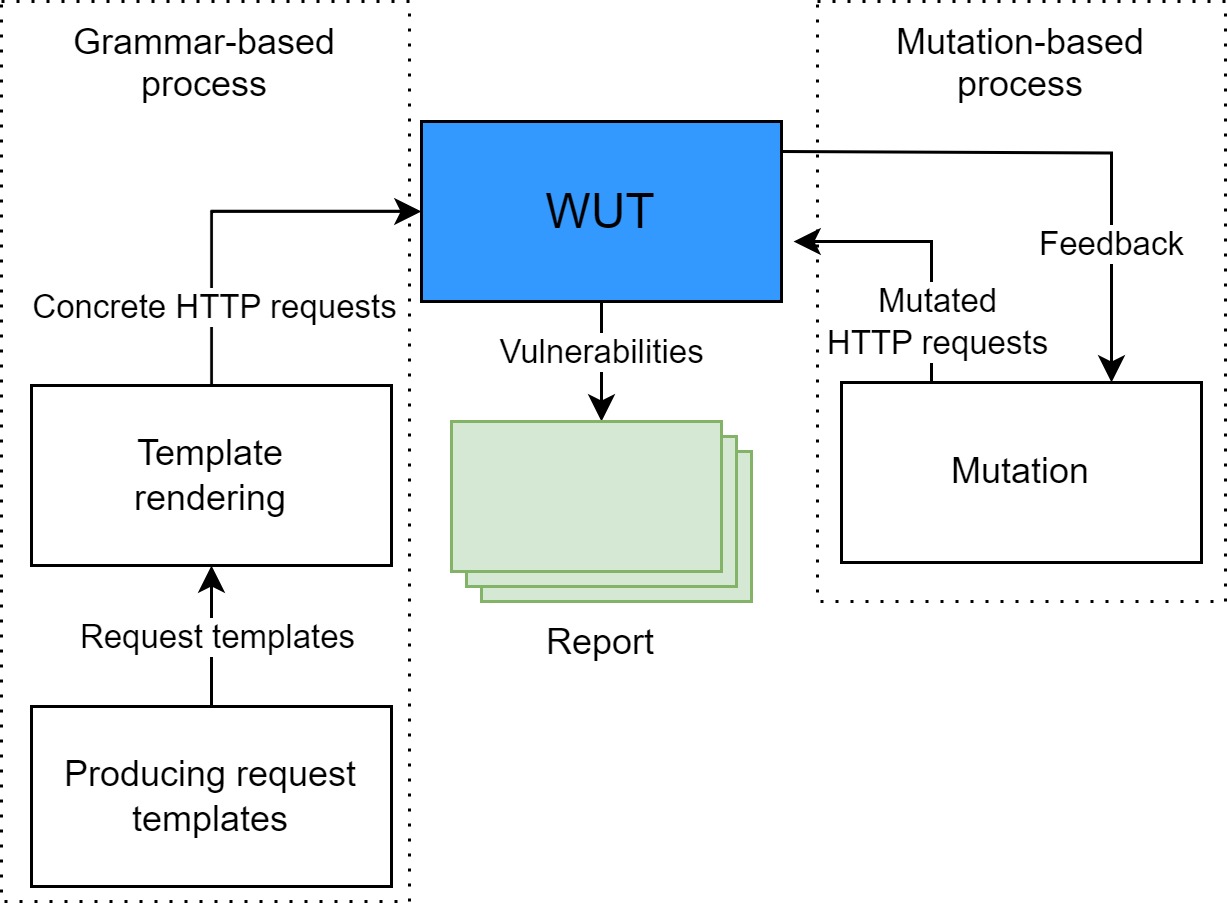}
    \caption{General overview of web API fuzzing framework. The first two steps in the left (explained in Section \ref{request-template-generation} and \ref{template-rendering}) are classified as grammar-based processes, while the next steps in the right (explained in Section \ref{execution-and-feedback} and \ref{mutation-strategies}) are mutation-based process.}
    \label{fig:general-overview}
\end{figure}

\begin{figure*}
    \includegraphics[width=1\linewidth]{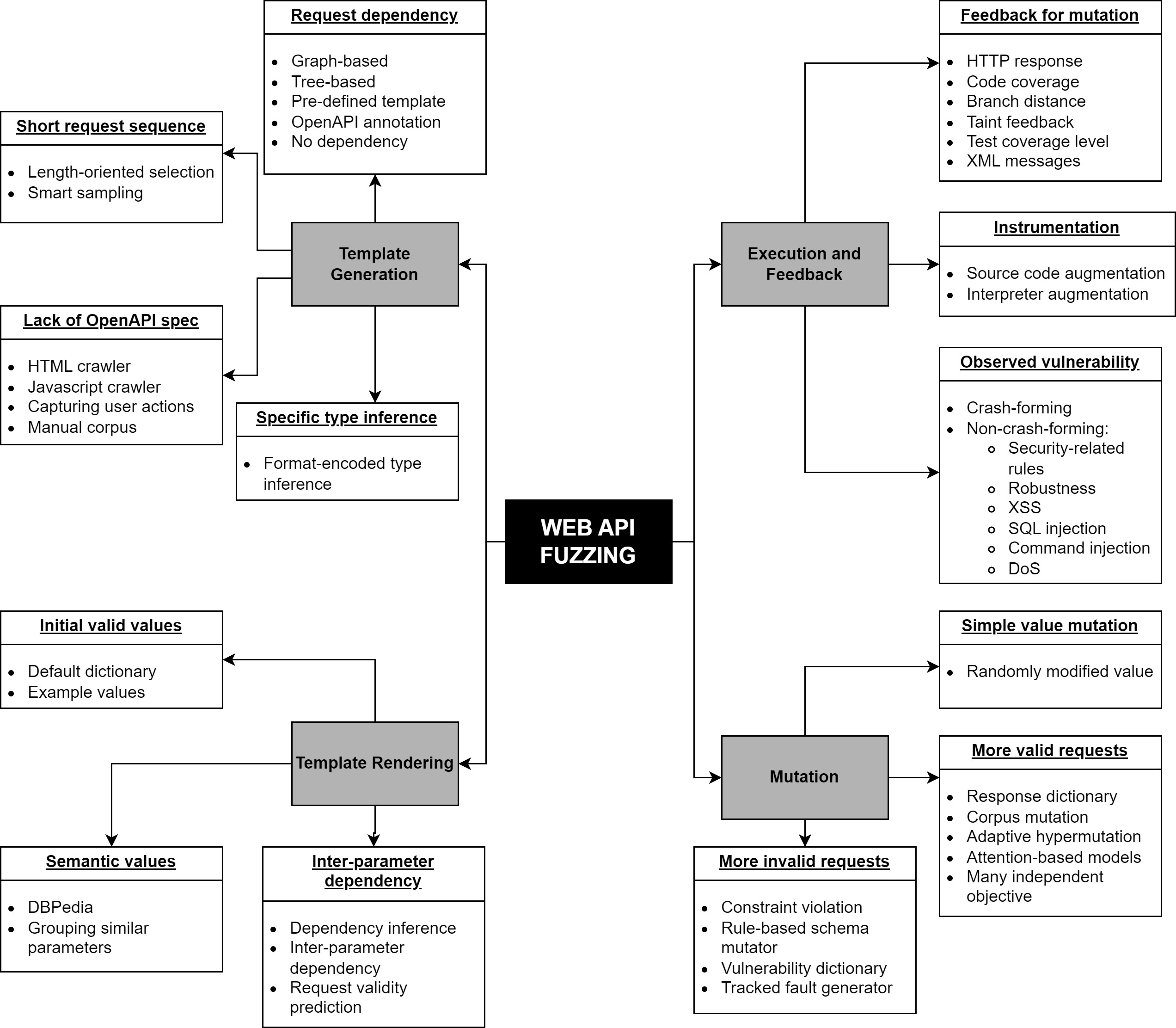}
    \caption{Taxonomy of the problems and solutions from prior web API fuzzing studies described in Section \ref{review-result-restful-api}. The grey boxes are the web API fuzzer processes explained from Sections \ref{request-template-generation} until \ref{mutation-strategies}. Each process has different problems along with their respective solutions (drawn in white boxes).}
    \label{fig:problem-solution}
\end{figure*}

In general, existing web API fuzzers involve four crucial processes: producing request templates, rendering them to produce concrete HTTP calls, executing them and getting feedback from WUT (Web Under Test) and mutation (illustrated in Figure~\ref{fig:general-overview}). The first two are called grammar-based processes because they produce completely new HTTP requests from scratch by employing grammar. Then the remainder processes are called mutation-based because they take existing HTTP requests and gradually alter them. We then classify the problems and solutions developed by existing studies into these processes and finally result in the problem-solution taxonomy (see Figure~\ref{fig:problem-solution}). All of those processes, along with the problems and solutions, are elaborated in the sections below.

\subsection{Request Template Generation}
\label{request-template-generation}
\noindent\fbox{%
    \parbox{\dimexpr\linewidth-2\fboxsep-2\fboxrule}{\textbf{RQ1}: What techniques are used to generate HTTP request templates?}%
}%
\\ \\
To produce initial test cases automatically, both standard and vulnerability-driven web API fuzzers first generate HTTP request templates. \textbf{HTTP request template generation} is the process of analysing given information to produce diverse request templates, which is essential in assembling long request sequences to reach deeper WUT statements and explore more states. This process can also be called \textbf{grammar generation} because it constructs the grammar sets that define the structure and format of the input data. Since creating templates is challenging considering the limited information that can be used as a reference, several studies have identified certain problems related to this issue and proposed some solutions as follows.

\subsubsection{Problem: Request dependency}
Most existing works raise a request dependency problem because it is unclear how to create correctly sequenced HTTP requests. Basically, the frameworks use the Open API specification as the guidance to produce valid HTTP requests. However, the document does not provide clear information regarding the order in which the HTTP requests are made --for example, which requests must be called first and which can be called later. When the request sequence is chaotic, the web server probably cannot process most requests because certain web states or conditions have not been met to execute the commands on those requests. For instance, the web server cannot execute a delete request if the resource to be deleted does not exist. Therefore, some studies developed diverse ways to solve this problem.

\begin{figure}
\centering
\includegraphics[width=0.7 \linewidth]{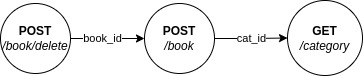}
\caption{An example of the dependency graph model. The \textit{get category} API will produce a cat ID that will be needed for calling the \textit{post book} API. The latter will generate a book ID used by the \textit{delete book} API.}
\label{pic:graph}
\end{figure}

\paragraph{Graph-based dependency}
Some researchers have developed a dependency graph to construct the relationship among HTTP requests. This graph provides clear guidance to find which requests must be committed before calling a specific request. The researchers proposed various techniques to build this graph, especially in deducing the request dependency. Atlidakis et al. \cite{atlidakis_restler_2019} infer the request dependencies by using the request types declared in the specification. RESTler, the framework they built, analyses the specification to determine which resources in one response become requirements in another request (see Figure \ref{pic:graph}). For example, calling a \textit{new item} request will produce an \textit{item id} in the response, which this \textit{id} field has to be inserted when calling an \textit{update item} request. Then, it can be concluded that the \textit{update item} request requires calling the \textit{new item} request first because of the \textit{id} field. Corradini et al. \cite{corradini_automated_2022}, other researchers who adopted the RESTler method, explained that identifying those dependency fields might be tricky because the field names in different operations are sometimes written differently, yet they are the same semantically. Therefore, those researchers use several strategies to match the field names, namely case sensitivity, ID completion, and stemming. Moreover, they utilise the CRUD semantics to determine the request order. For instance, the POST requests have to be called first before the GET or DELETE requests. Another researcher, Yamamoto \cite{yamamoto_efficient_2021}, designed a bipartite graph to ease the request dependency inference. The graph describes the relation between APIs and named values that appear in both request and response parameters.

\begin{figure}
\centering
\includegraphics[width=0.9 \linewidth]{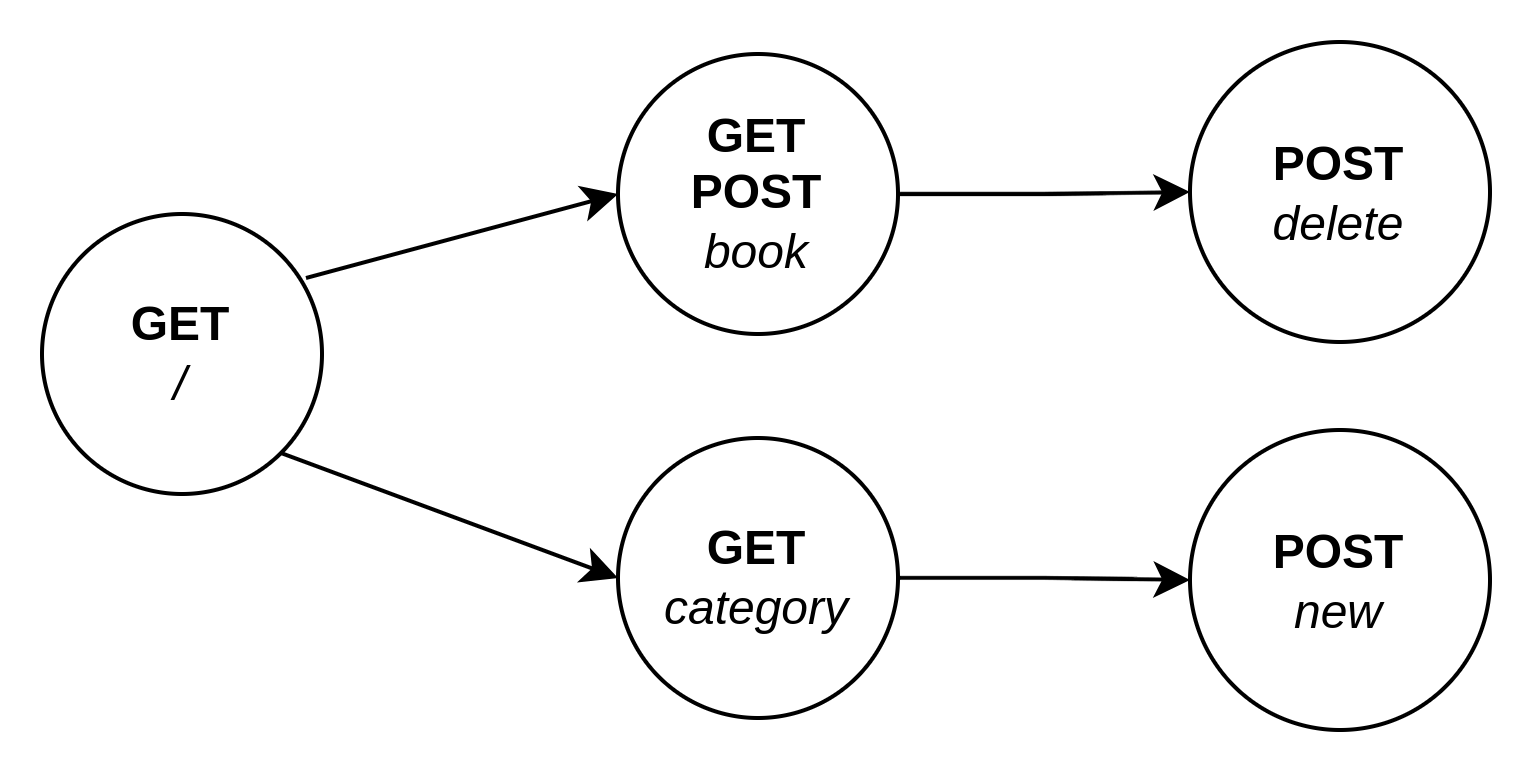}
\caption{An example of the tree model proposed by Lin et al. \cite{lin_forest_2023}. The tree describes valid endpoints that can be accessed: the web API root (\url{/}), \url{/book}, \url{/book/delete}, \url{/category}, and \url{/category/new} with some HTTP methods}
\label{tree-model}
\end{figure}

\paragraph{Tree-based dependency}
Lin et al. \cite{lin_forest_2023} found the classic dependency graph is inefficient because of the dependency explosion among APIs. This explosion led to too many possible paths that will overwhelm the fuzzing when traversing a complex graph model to explore APIs. As a result, they developed a tree-based dependency that is much simpler because the fuzzing frameworks only need to traverse a tree (which is linear complexity) instead of traversing a graph (which is quadratic complexity). This dependency is inspired by the fact that web users generally execute the parent node before calling its child nodes. A valid URL comprises several components separated by the slash (/) character. Nodes represent these components, and an edge between two nodes appears if a valid URL contains those components. Illustrated in Figure \ref{tree-model}, the tree model depicts the relationship between a root URL and its child URLs in which a node can contain one or more HTTP methods. Traversing this model via a depth-first or bread-first order can be more straightforward. In addition, Wu et al. \cite{wu_combinatorial_2022} use such hierarchical relations of resources that can be depicted in a tree structure to generate request sequences. Specifically, they proposed a constraint handler based on the hierarchical relations of resources and CRUD semantics to construct the template dynamically. The handler checks if each request in a template does not access a resource before its creation or after its deletion. For instance, the handler rejects the construction of a sequence consisting of GET /book/{id} and POST /book because it accesses the bookID before being created by the POST method.

\paragraph{Pre-defined template} 
The pre-defined template is a list or table containing the HTTP request order of an API. Tsai et al. \cite{tsai_rest_2021} prefer to use manual input combined with depth-first search to solve the API dependency and create the template. On the other hand, Zhang et al. \cite{zhang_resource-based_2019} proposed two templates: an independent template as an action that does not affect follow-up actions on any resource and a non-independent template as an action whose effects cannot be predetermined (see Table \ref{template-model-table}). Then, they also designed a smart sampling method (explained in Section \ref{smart-sampling}) to construct a request sequence based on both templates. 

\begin{table}
\caption{Some examples of the template model proposed by Zhang et al. \cite{zhang_resource-based_2019}}
\label{template-model-table}
\begin{tabular}{lp{3cm}ll}
 \hline
  & Description & Independent? & Template \\
 \hline
 1 & To retrieve a resource & YES & GET \\ 
 2 & To create a resource & NO & POST \\
 3 & To create an existing resource & NO & POST-POST \\
 \hline
\end{tabular}
\end{table}

\paragraph{OpenAPI annotation}
Instead of generating graph or tree-based dependency, a fuzzing framework can put the information, i.e., annotate the OpenAPI specification, to add more comprehensive information. Deng et al. \cite{deng_nautilus_nodate} use this idea to propose a set of customised annotations for OpenAPI Specification to help fuzzing generate desired request sequences. They introduced several types of annotations. For example, \texttt{dep-operation} expresses the dependent operations that must be performed before a certain operation. Since the annotations are human-readable and automatically processable, they can be performed by both human experts and fuzzing modules. Fuzzing modules infer the dependency based on the response fields of a certain operation that are used in other requests. 

\paragraph{No dependency}
Some web API fuzzing frameworks do not describe any specific model to infer valid consecutive API requests because the frameworks only construct one HTTP request for each test case. Some examples of existing works implementing this approach are the works of Van Rooij et al. \cite{van_rooij_webfuzz_2021}, and Trickel et al. \cite{trickel_toss_2023}. They did it because they mainly focused on enabling binary fuzzing tools or methods in web applications where the original tools in the binary applications do not require input sequences.

\subsubsection{Problem: Short request sequence}
Another problem mentioned by existing studies is the short request sequences. Since existing web API fuzzers tend to produce such sequences, Lyu et al. \cite{lyu_miner_nodate} stated that a long request sequence is essential to reach deeper states in web applications because it can cover many possible request combinations. One of these combinations may be valid for exploring statements hidden in hard-to-reach states. The authors also conducted a preliminary experiment showing more than 75\% of the issues in the GitLab services -the WUT- can only be reached using a long request sequence. Specifically, the authors took at least 3 HTTP requests in a sequence to reproduce those issues. Therefore, producing long request sequences can enable better testing coverage. Several methods can be used to produce longer request sequences, summarised below.

\paragraph{Length-oriented selection}
Lyu et al. \cite{lyu_miner_nodate} developed the length-oriented sequence construction module to construct candidate sequence templates. Initially, they employ the graph-based dependency as in the RESTler \cite{atlidakis_restler_2019} to produce initial sequence templates. Then, because the module uses a custom probability function, longer templates will have a higher chance of being chosen for the next stage: the extension process. This process puts a new request at the end of each chosen template. The newly generated templates will be retained if they bring valid responses. Therefore, each fuzzing iteration will produce longer and longer request sequences.

\paragraph{Smart sampling}
\label{smart-sampling}
Arcuri \cite{arcuri_restful_2019} proposed a smart sampling technique and pre-defined templates (explained in the prior section, see Table \ref{template-model-table}) to construct an individual test case containing some consecutive HTTP requests. The sampling involves four methods that can be chosen, namely by sampling: a resource with an independent template, a resource with a non-independent template, two resources, and more than two resources in which the last two only allow the last resource with non-parameter GET. Using this sampling technique will help fuzzing produce longer and longer sequences because other requests will be added before the existing request. The added requests placed in front are expected to put the WUT into the correct state for executing the existing request.

\subsubsection{Problem: Lack of OpenAPI specification}
Most web API fuzzing frameworks use OpenAPI specification documents to retrieve available API requests in the web application. Many web applications do not have OpenAPI specifications, but the fuzzing frameworks can rope with this by crawling and parsing the web client pages or employing a human.

\paragraph{HTML crawler}
Duchene et al. \cite{duchene_ligre_2013} proposed a state-aware crawler to parse the HTML documents and learn the control flow of the web application. The crawling process results in a control flow model (CFM) representing web pages (node) and requests (transition), and then the results are used by the KameleonFuzz \cite{duchene_kameleonfuzz_2014}, the authors' fuzzing framework. Van Rooij et al. \cite{van_rooij_webfuzz_2021} developed a similar HTML crawler to scan links from \textit{anchor} and \textit{form} elements in each HTML document reached using the \textit{html5lib} library. Lastly, this similar crawler workflow was also used by Witcher, the most recent fuzzing work developed by Trickel et al. \cite{trickel_toss_2023}. Given an entry point URL by the user, the Witcher's crawler works by scanning the HTML document responded to by the web server to find links and relevant fields that can establish HTTP requests, like form, input, select, and textarea elements.

\paragraph{Javascript crawler}
This source may be the best to crawl the URLs compared to HTML-formed responses since modern web applications often put their URLs to the web server in JavaScript documents. As an alternative to parsing the HTML documents, Gauthier et al. \cite{gauthier_experience_2022} developed BackREST using a client-side javascript document to obtain a REST API inference model. They demonstrated one example: the web API entry points in a javascript document used in the Node.js Express application are similar to those in the Open-API Specification. Since these documents reveal similar information, BackREST introduced a state-aware crawler to analyse the javascript code that calls web APIs. The crawling process yields valid HTTP requests, and the requests executed by the browser are intercepted by the Man-In-The-Middle (MITM) proxy to build an API inference model. This similar approach was also adopted by Yandrapally et al. \cite{yandrapally_carving_2023}, who designed an API test carving to produce an API-level test suite and a test execution report and OpenAPI specification. Their work is intended for all web applications, irrespective of the web frameworks they use. 

\paragraph{Capturing user actions}
Instead of employing crawlers to obtain valid APIs, Fung et al. \cite{fung_scanning_2014} utilise a human to interact with the WUT in the browser as usual. They built a browser add-on to capture the user actions, and the add-on can track the parameter dependency from the submission requests and the server responses. Finally, the capturing phase will produce valid request templates that can be mutated later. 

\paragraph{Manual corpus}
Rather than asking users to do some actions on the WUT, some papers prefer to use the user's initial input (i.e., corpus) for the fuzzing frameworks. This is the same as the binary fuzzing frameworks that assume the users are experts who can provide valid inputs that match the requested format. Woo et al. \cite{woo_re-checker_2018} design a web API fuzzing framework that receives a seed file containing a default HTTP header and payload based on a JSON format. The framework then mutates the input contained in the file to test the WUT.

\subsubsection{Problem: Specific type inference}
The previous works on web API fuzzing have not addressed the root cause of request candidate space explosion, which is the limited data types available to represent API parameters. For example, a study by Lei et al. \cite{lei_bootstrapping_2023} showed that most web APIs use string-typed parameters, and the string type has a huge space to be explored. Generating effective string values that can pass parameter validation, reach business logic, and trigger bugs is challenging because there is little information to narrow down the input space of the parameter. Therefore, to address this issue, recent work focused on inferring the information about the data types, as discussed below.

\paragraph{Format-encoded Type (FET) inference}
Lei et al. \cite{lei_bootstrapping_2023} proposed a FET inference technique to provide a fine-grained description method for parameters that utilises data type and value format. FET can be defined as new data types that are more specific than the conventional types the software developers are using in programming (e.g., string or integer). To construct the FET lattice, the authors referred to popular RESTful services (a total of 1268 APIs). They resulted in 21 ubiquitous FETs organised in 5 levels. All API parameter values are classified into those FETs to enable the application of different mutation strategies. For example, the FET of datetime (value example: "2019-2-29") has a different treatment from the FET of hash (value example: "19CGHEE2") even though both are string-typed. Fuzzing will apply a strategy to make year overflows in the datetime parameter and to use a non-hexadecimal-number string in the hash parameter.
\\

\noindent\fbox{%
    \parbox{\dimexpr\linewidth-2\fboxsep-2\fboxrule}{\textbf{Answer to RQ1}: To generate HTTP request templates, the most prominent solutions are: 1) creating request dependency (either graph-based or tree-based) inferred from OpenAPI specification, and 2) using pre-defined templates. The others are OpenAPI annotation, length-oriented selection, smart sampling, HTML/JS crawlers, capturing user actions, manual corpus, and format-encoded type inference.}%
}%

\subsection{Template Rendering}
\label{template-rendering}

\begin{figure*}
\includegraphics[width=1 \linewidth]{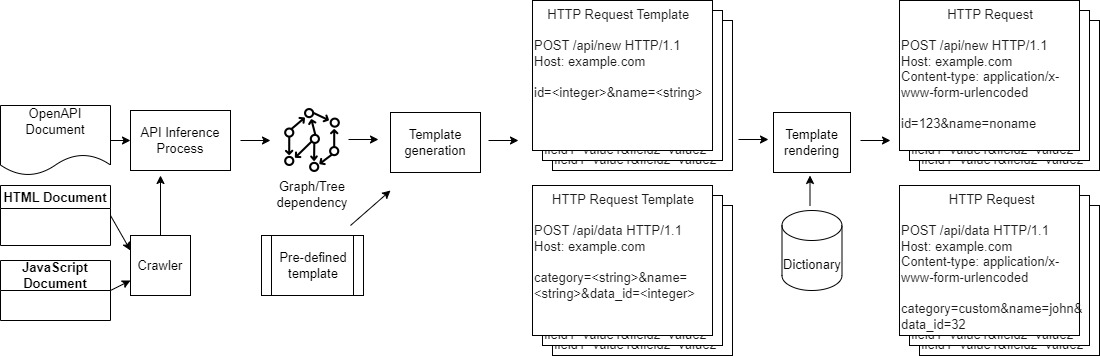}
\caption{Illustration of generating concrete request sequences described in sections \ref{request-template-generation} and \ref{template-rendering}.}
\label{fig:request-generation}
\end{figure*}

\noindent\fbox{%
    \parbox{\dimexpr\linewidth-2\fboxsep-2\fboxrule}{\textbf{RQ2}: How are the request templates rendered?}%
}%
\\ \\
After generating request template sequences from the prior step, existing web API fuzzing frameworks commonly continue to concretise the sequences. This process is known as \textbf{template rendering}, which takes generic input templates and populates them with concrete values to create real requests that can be sent to the WUT (see Figure \ref{fig:request-generation}). These two initial processes (request template generation and template rendering) can also be called \textbf{grammar-based input generation} because the input templates define the structure and format of the input data, and fuzzer will fill in the placeholders in the templates. The processes allow the fuzzing framework to generate valid inputs that conform to the expected format of the WUT. Several problems arise in this stage, as follows.

\subsubsection{Problem: Initial valid values}
\label{valid-values-problem}
Fuzzer will miss opportunities to test deeper web functionalities if it produces initial HTTP requests that do not comply with the WUT's requirements. The WUT will reject those malformed requests, and then the mutation process struggles to modify them to be valid. In the end, no matter how long the fuzzing process operates, it is very likely that fuzzing will fail to explore interesting features in the WUT. Therefore, initial valid values serve as a good starting point for the fuzzing campaign. However, some studies mentioned the complexity of finding valid values because the OpenAPI specification rarely provides valid examples of each API field. This situation made the researchers design a default dictionary in the template rendering process.

\paragraph{Default dictionary and example values (DD)}
Basically, all existing web API fuzzing frameworks prepare default values for each data type to concretise the request templates. Atlidakis et al. \cite{atlidakis_restler_2019} use a user-configurable value dictionary to fill the request templates based on the field types. For example, they set values 0, 1, and -10 for fields requesting integer values and "sampleString" for fields requesting string values. Other examples are Godefroid et al. \cite{godefroid_intelligent_2020} and Lyu et al. \cite{lyu_miner_nodate}, which defined various valid value options for each data type. Besides using this technique, other researchers also found that some OpenAPI specifications may provide the \textit{example} fields that hold concrete example values documented by the web developers. Therefore, apart from using their default values, some researchers, such as Corradini et al. \cite{corradini_automated_2022} and Deng et al. \cite{deng_nautilus_nodate}, utilise the \textit{example} field to fill the initial values of HTTP request templates.

\subsubsection{Problem: Semantic value}
Some works mention semantic value problem, which means the fuzzers have to fill in values that are not only syntactically valid but also semantically valid. The main reason is that the human testers are usually obliged to give meaningful, valid inputs for each API parameter \cite{alonso_arte_2023}. In real-world API applications, providing only syntactically valid but semantically invalid inputs will bring unsatisfactory results. For example, the search operation in the YouTube API application demands the user to fill in semantically valid values for every search field. If not fulfilled, the application will not show a result because it does not match with any data. This means the user cannot thoroughly test the search feature without the semantically valid values. In addition, producing semantically valid inputs is challenging because no semantic information can be derived from the OpenAPI specification. Therefore, some studies proposed a strategy to address this issue as follows.

\paragraph{DBPedia}
\label{dbpedia}
Previous works used external knowledge dictionaries to help fuzzing. Alonso et al. \cite{alonso_automated_2021} \cite{alonso_arte_2023} leverage semantic knowledge discovery to generate realistic test inputs. They used DBPedia \cite{bizer_dbpedia_2009}, a knowledge base, because they had identified its potential, but no one had implemented it in the web testing domain. However, employing this dictionary is not straightforward because of its diverse domains. For example, when fuzzing demands the concrete value for a \textit{title} field, the DBPedia will return hundreds or even thousands of diverse title-type inputs (e.g., movies, games, or books) \cite{alonso_arte_2023}. The authors managed to narrow down the results by inserting more specific criteria derived from HTTP parameter combinations. For instance, adding the \textit{publisher-name} field in the search queries with makes the results more specific to book title values.

\paragraph{Grouping similar parameters}
Liu et al. \cite{liu_automated_2023} group the string parameters with the same semantic meaning to generate semantically correct results. For example, \textit{loginid}, \textit{username}, and \textit{account} have a similar semantic meaning that demands the user to provide a name. Therefore, the grouping module will utilise certain modules to produce correct names. Another example is the group of the description, message, and comment field that requires the user to give a description text.

\subsubsection{Problem: Inter-parameter dependency}
\label{inter-parameter-dependency-problem}
Inter-parameter dependency means that one or several parameter values depend on other parameters. For example, the Youtube API documentation states that the \textit{type} parameter must be filled with \textit{video} before setting up the \textit{videoDefinition} value \cite{yangui_catalogue_2019}. Therefore, if the \textit{type} parameter is not filled in, the web server will not process the value of the \textit{videoDefinition} parameter. Another example is that a parameter representing the marital \textit{status} of an individual must have the value of \textit{married} before filling in the \textit{spouse-name} field. Martin-Lopez et al. \cite{yangui_catalogue_2019} raised this issue because OpenAPI specification does not inform this dependency explicitly. Their study revealed that around 85\% of APIs from 40 real-world applications (containing more than 2.5K APIs) have such parameter dependency rule. Without satisfying the rule, a fuzzing campaign cannot go further to test the corresponding functions. To overcome this problem, some studies proposed solutions as follows.

\paragraph{Dependency inference}
Wu et al. \cite{wu_combinatorial_2022} employ a Natural Language Processing (NLP) method to infer the parameter dependency constraints that are usually written informally in the \textit{description} field of the Open-API Specification documents. They leverage spaCy\footnote{https://spacy.io/}, an open-source library, to perform this task. Before a fuzzing campaign begins, they prepare some patterns to look for, such as \textit{if PARAM\_A is VALUE\_B, PARAM\_C is required}. Then, using the pattern-based matching engine, the library will check if any text in the Open-API Specification documents matches those patterns. Adopting a similar idea, Kim et al. \cite{kim_enhancing_2023} designed an NLP-based rule extraction to infer the rule and inter-parameter dependencies from the description fields. The extraction process involves vocabulary term identification, value and parameter name detection, and rule generation. Finally, the process produces OpenAPI-compliant rules that are machine-readable.

\paragraph{Inter-parameter Dependency Language (IDL)}
Martin-Lopez et al. \cite{martin-lopez_specification_2022} presented a domain-specific language called IDL (Inter-parameter Dependency Language) to express seven types of inter-parameter dependencies: \textit{Requires, Or, OnlyOne, AllOrNone, ZeroOrOne, Arithmetic/Relational}, and \textit{Complex}. In this language, those dependencies are expressed using invariants, conditional definitions, logical operators, and relational operators. However, this study still uses manual work to create the corresponding IDL from the API documentation. Apart from IDL, the study also proposed an automated analysis tool to check whether a request meets all the dependency constraints. Ultimately, the tool and IDL can automatically generate concrete requests satisfying the inter-parameter dependencies \cite{martin-lopez_restest_2021}. 

\paragraph{Request validity prediction}
Mirabella et al. \cite{mirabella_deep_2021} employ a deep learning model to predict if a certain HTTP request is valid that satisfies all the input constraints, including the inter-parameter dependencies. The training data is a dataset of API calls with valid or invalid labels. The learning model consists of the input layer that fits the data frame numbers, five inner layers (32, 16, 8, 4, and 2 neurons, respectively), and the output layer (one neuron). If the output layer produces a value greater than 0.5, the input is valid; otherwise, it is faulty.
\\

\noindent\fbox{%
    \parbox{\dimexpr\linewidth-2\fboxsep-2\fboxrule}{\textbf{Answer to RQ2}: To render the request templates, the most prominent solutions are using default dictionary and example values. The others are initial valid values, semantic values, inter-parameter dependency, DBPedia, grouping similar parameters, dependency inference method, inter-parameter dependency language, and request validity prediction.}%
}%

\subsection{Execution and Getting Feedback}
\label{execution-and-feedback}
Fuzzer sends all concrete HTTP requests produced by the prior steps to the WUT and then receives the reply. Based on the reply, the fuzzer will decide which requests or sequences must be explored and what mutation strategies must be applied. In this stage, the fuzzing researchers mention several problems related to reply messages, as follows.

\subsubsection{Problem: Feedback for mutation}
\noindent\fbox{%
    \parbox{\dimexpr\linewidth-2\fboxsep-2\fboxrule}{\textbf{RQ3}: What kind of feedback is extracted from the WUT?}%
}%
\\ \\
Feedback from the WUT is crucial for the mutation process because it helps comprehend the impact of the generated requests. If supplied with helpful feedback, fuzzer will work better because it can neglect the ineffective requests to exploit the effective ones. In addition, particular feedback helps fuzzer to identify areas of the code that have not been tested and generate inputs specifically to explore those regions. This can increase the likelihood of finding hidden vulnerabilities. Prior web API fuzzing studies observed the following feedback information.

\paragraph{HTTP response}
HTTP response is the standard information a web application provides after executing an HTTP request. Since it is common feedback, almost all web API fuzzing frameworks utilise this information. The response generally consists of an HTTP response code and message (as explained in Section \ref{http-method-response-code}). Several examples of existing fuzzing frameworks utilising this information are as follows. A fuzzing framework built by Lyu et al. \cite{lyu_miner_nodate}, MINER, re-uses request sequences that bring valid HTTP response codes to produce longer sequences. Similarly, Wu et al. \cite{wu_combinatorial_2022} developed a fuzzer called RestCT that obtains successful requests to utilise several constrained covering arrays for generating next-round requests. Then, a fuzzing framework called ResTestGen built by Corradini et al. \cite{corradini_automated_2023} collects requests with successful status codes to be used to create new test cases.

\paragraph{Code coverage}
\label{code-coverage}
Code coverage is how much code in a WUT is executed when the WUT receives inputs. This information is really powerful for fuzzer to keep track of its testing progress because the WUT generally consists of many conditional statements, making some codes probably unexecuted. Given code coverage is not standard feedback in web applications, it can only be enabled by implementing instrumentation, which will be provided in Section \ref{instrumentation-problem}.
Prior works employed code coverage to enhance web API fuzzing frameworks. Van Rooij et al. \cite{van_rooij_webfuzz_2021} use coverage score and other metrics to rank requests. The AFL algorithm  \cite{noauthor_technical_nodate} highly inspired their work to keep a new high-score request for future mutation. The algorithm also imbues Trickel et al. \cite{trickel_toss_2023} to develop a web API fuzzer called Witcher. This work augmented the web interpreter to produce code coverage so that the Witcher could store and mutate the request that brought a new execution path.

\paragraph{Branch distance}
\label{branch-distance}
To get more precise information on which conditions need to be solved to go deeper into the code, Arcuri \cite{arcuri_evomaster_2018} calculates a branch distance, which is how close a particular input is to solving the constraints \cite{korel_automated_1990}. For example, the conditional branch of \textit{if (a==20)} is reached by an HTTP request with parameter \textit{a=15}. Therefore, the branch distance of the request is 5. To be able to calculate the distance, custom instrumentation is needed. In some cases, satisfying the conditional branches may be complicated because the involved variables are not user-controlled data that are not required in the HTTP request. Instead, those are influenced by other functions. Therefore, the fuzzing framework proposed by Arcuri \cite{arcuri_evomaster_2018} tends to pick requests with low branch distances for the mutation stage because those are relatively close to solving the branch.

\paragraph{Taint feedback}
Taint analysis and tracking are important techniques in fuzzing because they provide information about how data flows through a program, highlighting the influence of external input on the program's state \cite{tripp_taj_nodate}. Taint feedback, the result from the taint analysis, helps to understand the input data propagation through the program. Marking certain input data as tainted can trace the input flow and identify potential security-sensitive operations that involve user-controlled data. Using this approach, the fuzzing framework can focus on generating inputs specifically targeting the paths and operations influenced by HTTP requests to increase the chances of discovering vulnerabilities. Gauthier et al. \cite{gauthier_experience_2022} proposed a taint-driven fuzzing containing taint analysis that is executed in the initial stage before starting the fuzzing campaign. During the campaign, the analysis will result in taint feedback on which requests reach security-sensitive program locations that are already defined. Arcuri et al. \cite{arcuri_enhancing_2022} use the taint analysis to track variables at runtime for seeding strategies.

\paragraph{Test Coverage Level (TCL)}
Developed by Martin-Lopes et al. \cite{martin-lopez_test_2019}, TCL assesses the coverage of the request collections. It roughly measures the extent to which the WUT's code has been exercised by the generated test inputs using the black-box approach. It uses input and output criteria (e.g., paths, operations, and content type) that can be observed without having the source code. TCL0 represents the weakest coverage level, and the strongest one is TCL7. Each TCL level has different criteria that the requests must satisfy. To reach a certain TCL, the request collection must meet all requirements belonging to the previous levels. Tsai et al. \cite{tsai_rest_2021} developed a fuzzing framework called \textit{HsuanFuzz} employing this TCL concept. If request collection can increase the TCL of a certain API path, it will be stored in the fuzzing corpus. 

\paragraph{XML messages}
Jan et al. \cite{jan_automatic_2019} designed a fuzzing framework to exploit XML injection vulnerabilities in WUT. To know how close the generated HTTP requests are to the XML injection, they compare the XML messages produced by the WUT and a set of well-formed yet malicious XML messages that they call TO (Test Objective). Since not all WUTs produce XML messages, they determined their work to test the "front-end" web application of SOA (Service-oriented Architecture) systems. The front end mentioned here does not refer to the client-side web application that runs in the browser (e.g., Javascript) but points to a server-side web application that acts as the front gate for the diverse back-end systems in SOA systems.
\\

\noindent\fbox{%
    \parbox{\dimexpr\linewidth-2\fboxsep-2\fboxrule}{\textbf{Answer to RQ3}: The feedback that is most often extracted from the WUT is the HTTP response code. The others are code coverage, branch distance, taint feedback, TCL, and XML messages.}%
}%

\subsubsection{Problem: Instrumentation}
\label{instrumentation-problem}
\noindent\fbox{%
    \parbox{\dimexpr\linewidth-2\fboxsep-2\fboxrule}{\textbf{RQ4}: How are the WUTs instrumented?}%
    % {\textbf{RQ4}: What kind of instrumentation methods are used for the WUT?}%
}%
\\ \\
Code instrumentation means an instrumentation tool will put some probes in the WUT's source code to collect how much code the WUT reaches during the execution \cite{horvath_code_2019}. The instrumentation can be done either in a static or dynamic way, in which the former means it is performed at compilation time and has less run-time overhead; meanwhile, the latter is done at run-time and has better performance \cite{manes_art_2021}. Since the black-box web API fuzzers do not have access to the source code, they cannot do this process; otherwise, grey-box and white-box fuzzers can. Even though the instrumentation is helpful for fuzzers, it is only employed by a few prior fuzzing frameworks because its implementation is challenging. Web applications that operate in various platforms may force the fuzzer developers to develop a new instrumentation tool. The following are the instrumentation techniques employed by prior web API fuzzers to instrument the WUT.

\paragraph{Source code augmentation}
\label{library-customisation}
Putting probes in the source code or byte code level is the common way to instrument applications, in which around 28\% of prior studies adopt this technique.
EvoMaster \cite{arcuri_evomaster_2018} developed new instrumentation tools to get code coverage from various WUT platforms. First, for Java-based WUT, Arcuri et al. \cite{arcuri_restful_2017} instantiate a Java agent to intercept all class loadings and add probes in the bytecode. On the other hand, for Javascript-based WUT, Zhang et al. \cite{zhang_javascript_2023} develop a plugin for Babel (a JavaScript transpiler) to create an instrumented version of the WUT which contains probes in the source code. Rather than employing ready-to-use third-party libraries (e.g., JaCoCo), designing these custom tools enables EvoMaster to add extra features (e.g., calculating the branch distance) for the mutation feedback. Figure \ref{fig:instrumentation-example} illustrates the instrumented version of the code.

\begin{figure}
     \centering
     \begin{subfigure}[b]{0.5\textwidth}
         \centering
\begin{lstlisting}[language=PHP, firstline=1, lastline=2, numbers=left]
let x = 0;
\end{lstlisting}
         \caption{Original code}
         \label{fig:original-code}
     \end{subfigure}
     \hfill
     \begin{subfigure}[b]{0.5\textwidth}
         \centering
\begin{lstlisting}[language=PHP, firstline=3, lastline=7, numbers=left]
const __EM__ = require("evomaster-client-js").InjectedFunctions;
__EM__.registerTargets(["File_test.ts","Line_test.ts_00001","Statement_test.ts_00001_0"]);
__EM__.enteringStatement("test.ts",1,0);
let x = 0;
__EM__.completedStatement("test.ts",1,0);
\end{lstlisting}
         \caption{Instrumented version}
         \label{fig:instrumented-code}
     \end{subfigure}
        \caption{An example of the instrumentation process taken from the paper of Zhang et al. \cite{zhang_javascript_2023}, in which \ref{fig:original-code} is the original code written in typescript and \ref{fig:instrumented-code} is the instrumented version.}
        \label{fig:instrumentation-example}
\end{figure}

Van Rooij et al. \cite{van_rooij_webfuzz_2021} created an instrumentation method to get live information from PHP-based WUT. It instruments the application in the AST (Abstract Syntax Tree) to catch basic block or branch coverage. AST is a tree structure representing source code syntax without showing the details that can be used to identify statements or declarations in the program \cite{zhang_novel_2019}. Using this approach, this instrumentation method parses each file using the PHP-Parser library and identifies basic blocks during the AST traversal process. Then, it puts probes at the beginning of a basic block, such as the first statement in a function definition or a control function.

\paragraph{Interpreter augmentation}
Instead of augmenting the source code as explained above, Trickel et al. \cite{trickel_toss_2023} prefer to augment the interpreter application. In the context of web applications, an interpreter is a component running in the web server (recall Figure \ref{fig:web-architecture}) that reads the source codes of the server-side web application and translates them into byte-code instructions without the need for a separate compilation step. Their work augmented the interpreter by modifying and recompiling its source codes. The augmented interpreter will then call the proposed library function, Witcher's Coverage Accountant, to send the line number, opcode, and parameter of the current byte-code instruction during the WUT execution. The Accountant function aims to measure the test coverage from the sent data.\\

\noindent\fbox{%
    \parbox{\dimexpr\linewidth-2\fboxsep-2\fboxrule}{\textbf{Answer to RQ4}: Some existing web API fuzzer studies (28\%) use source code or interpreter augmentation (2\%) for the instrumentation. The others (70\%) do not instrument the WUT because they use the black-box testing approach.}%
}%

\subsubsection{Problem: Observed vulnerability}
\noindent\fbox{%
    \parbox{\dimexpr\linewidth-2\fboxsep-2\fboxrule}{\textbf{RQ5}: What vulnerabilities are observed?}%
}%
\\ \\
WUT generally exposes its vulnerabilities when receiving a request raising inevitable errors. Determining vulnerabilities to look for is a crucial issue because it highly influences the fuzzing framework design. Basically, the vulnerabilities observed by the existing web API fuzzers can be grouped into two big categories: %general crash and specific vulnerabilities.
crash-forming and non-crash-forming vulnerabilities.

% \paragraph{WUT crash}
\paragraph{Crash-forming vulnerabilities}
Standard fuzzers (recall Section~\ref{standard-fuzzing}) generate inputs to make software crash and look for any vulnerability behind the crash. The vulnerabilities found from in this way can be called crash-forming vulnerabilities. Adopting standard fuzzing workflow, most of the existing web API fuzzing frameworks (72\% of existing papers) report the WUT crash (Internal server error) and use HTTP status codes to infer to locate the vulnerability. 
%One notable point of the existing web API fuzzing frameworks is that most report the WUT crash (Internal server error) and use HTTP status codes to infer whether the crash is happening. 
For example, EvoMaster \cite{arcuri_evomaster_2018}, RESTler \cite{atlidakis_restler_2019}, RestTestGen \cite{viglianisi_resttestgen_2020}, RESTest \cite{kafeza_restest_2020}, foREST \cite{lin_forest_2023}, and RestCT \cite{wu_combinatorial_2022}, utilise this technique to know the success of the injected inputs by noting how many HTTP requests triggered 500-response codes. Then, the fuzzer users must check internal error information (e.g., full stack traces of thrown exceptions) to know what actual error is raised. However, to detect more vulnerabilities, the users cannot rely solely on this information because most web vulnerabilities do not manifest themselves as error status codes. Another issue with relying solely on error codes is a 5xx-response code does not always mean a software fault if the web API is connected to external services \cite{marculescu_faults_2022}. For example, a fuzzer sends a request to the first web service that relies on the second service. If the second is down, the first will return a 5xx-response code even though the first service runs normally without error. %Therefore, although the HTTP response code is essential to show that a problem occurred on the WUT, the fuzzer users need additional information to detect other vulnerabilities.

% \paragraph{Specific vulnerability}
\paragraph{Non-crash-forming vulnerabilities}
\label{specific-vulnerability}
In addition to the crash, to ensure malicious parties cannot exploit the WUT, some web API fuzzers catch specific vulnerabilities that do not form a crash. To detect these vulnerabilities, web fuzzers must be equipped with typical bug catchers, as follows. To improve the RESTler capabilities \cite{atlidakis_restler_2019} for capturing specific bugs (resource violation), Atlidakis et al. \cite{atlidakis_checking_2020} designed four security rules, namely the use-after-free rule, resource-leak rule, resource-hierarchy rule, and user-namespace rule. These rules are checked after getting the web responses. Take the second rule as an example. If a child resource of a parent resource can be accessed from another parent resource, it means the response breaks the resource-leak rule and is classified as a bug. Corradini et al. \cite{corradini_automated_2023} improved the RestTestGen's capability \cite{viglianisi_resttestgen_2020} in catching mass assignment vulnerabilities that can happen when external users can manipulate the value of a resource meant to be read-only by exploiting a misconfiguration of the automatic parameter binding. The authors proposed observing the WUT by checking whether the read-only attributes can be overwritten and whether they differ from their default values.

%Laranjeiro et al. \cite{laranjeiro_black_2021} defined the fault model containing 57 common faults in the web API to serve as a mutation rule. After the requests are injected into the web under test, bBOXRT analyses and classifies the service behaviour with a failure model scale: Catastrophic, Restart, Abort, Silent, and Hindering. 
Another web fuzzer, BackREST, was designed to catch web vulnerabilities like XSS, SQL injection, Command injection, and Denial of Service (DoS). It also enhanced its work with taint feedback to detect more SQL and Command injection vulnerabilities. Van Rooij et al. \cite{van_rooij_webfuzz_2021} designed a fuzzer to catch stored and reflective XSS vulnerabilities. Initially, their fuzzer injects XSS payloads in the HTTP request parameters, which will then call the alert function. Parsing Javascript code in the HTML responses using \textit{esprima}\footnote{https://github.com/Kronuz/esprima-python} library, WebFuzz --their fuzzer's name-- will check the corresponding alert function call. If affirmative of containing the alert, it can be concluded that the web application is vulnerable to XSS. In addition to WebFuzz, other web fuzzers used a similar approach to detect XSS bugs, such as State-Aware Vulnerability Scanner \cite{doupe1_enemy_2012}, KameleonFuzz \cite{duchene_ligre_2013}, Cefuzz \cite{zhao_cefuzz_2022}, and ZokFuzz \cite{zhang_zokfuzz_2022}.\\

\noindent\fbox{%
    \parbox{\dimexpr\linewidth-2\fboxsep-2\fboxrule}{\textbf{Answer to RQ5}: Most web API fuzzer studies (72\%) observe crash-forming vulnerabilities using the 500-response code. The others look for specific web vulnerabilities, namely the violation of certain security-related rules, XSS, SQL injection, and command injection.}%
}%

\subsection{Mutation}
\label{mutation-strategies}
\noindent\fbox{%
    \parbox{\dimexpr\linewidth-2\fboxsep-2\fboxrule}{\textbf{RQ6}: How are the existing input spaces expanded?}%
    % {\textbf{RQ6}: What techniques are used to expand the existing input space?}%
}%
\\ \\
The mutation process is the core of the fuzzing method because it gradually expands the existing input to find more relevant input space. Running a fuzzer for a long time will allow the mutator --the fuzzing's component doing the mutation process-- to generate more inputs to explore execution paths that contain uncovered vulnerable code. However, suppose the mutator is poorly designed and does not suit the characteristics of the targeted application. In that case, no matter how long the fuzzing continues, its effectiveness will remain low (e.g., code coverage will be flat or no bugs will be found). The following are several problems related to mutator design and solutions provided by previous works.

\subsubsection{Problem: Simple value mutation}
\label{simple-value-mutation}
Simple value mutation is a basic technique to alter or completely change the previous input values in a fuzzing campaign. Before other techniques appear, most fuzzing studies make simple modifications, such as random bit flipping or byte deletion, to gradually discover unexpected behaviour in the applications. However, those simple mutation techniques are not enough in the web application context because they were commonly intended for finding memory-related vulnerabilities in binary applications. Therefore, existing web API fuzzing proposals devise another way to do simple mutation in order to make the mutation process more suitable for web domains, as follows.

\paragraph{Randomly modified values (RMV)}
Creating random data may be the most straightforward technique for mutation in all fuzzing types, including web API fuzzing. However, web API fuzzer rarely applies mutation at the bit level. The work of Lei et al. \cite{lei_bootstrapping_2023} showed that most RESTful web applications use around 67\% string-typed and 32\% number-typed parameters. Therefore, all web API fuzzers create new mutated HTTP requests using string modification (e.g., duplicate random string or swap string position) or number operation techniques (e.g., multiplication) to fill the parameter values.

\subsubsection{Problem: More valid requests}
Since the simple value mutation can result in either valid or invalid values, the fuzzing framework should be careful when using that technique. Aggressively using a random generator may damage the data structure, leading to the rejection by WUT \cite{van_rooij_webfuzz_2021}; however, data changes that are too small may not be enough to trigger new execution flows. Therefore, based on the objectives, previous studies designed the mutation process only to generate either valid or invalid requests. Those who want WUTs to execute all statements behind branch conditions must send more valid requests that can satisfy the conditions. To achieve that goal, they have designed the following strategies.

\paragraph{Response dictionary (RD)}
\label{response-dictionary}
Besides using random data, Viglianisi et al. \cite{viglianisi_resttestgen_2020} use a response dictionary containing a map between field names and their valid values to complement the default dictionary. The valid values come from the values that appear in the valid HTTP responses. Reusing the already tested values is effective, especially for certain types whose values are generated by the WUT, such as id. However, because of the developer's carelessness, the responses that are supposed to use the same parameter names may bring names that are slightly different to the existing ones yet are the same semantically. For instance, a response from the WUT contains \textit{nameid}, but another response from a different function carries \textit{id\_name} instead of \textit{nameid}. Hence, the authors also prepared strategies to match similar field names. They match those names with other variations that may happen (e.g., \textit{nameid} should be matched to \textit{id\_name}). Lin et al. \cite{lin_forest_2023} also use the response dictionary idea to build a hierarchically tree-shaped resource pool to hold each resource's possible values. This pool data is extracted from response messages.

\paragraph{Corpus mutation (CM)}
In the context of web API fuzzing, a corpus is the place to put interesting HTTP requests (e.g., a request that brings a crash). Some researchers maintain this corpus to reuse such requests in the mutation process. For example, Trickel et al. \cite{trickel_toss_2023} blend the combination of parameter names and values between the requests stored in the corpus to generate a mutated request.

\begin{figure*}
    \centering
    \includegraphics[width=0.6\linewidth]{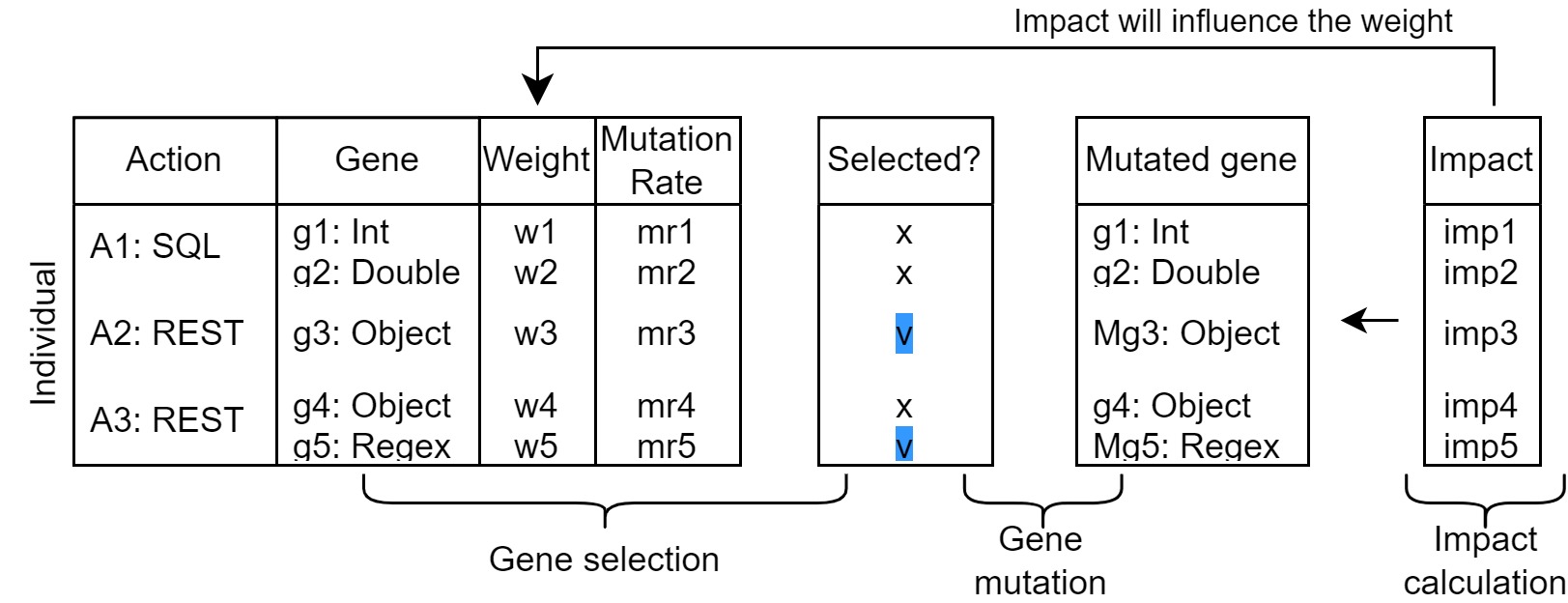}
    \caption{Illustration of the adaptive hypermutation proposed by Zhang et al. \cite{zhang_adaptive_2022}. One individual consisting of some actions (e.g., HTTP requests) has many genes, and each gen has a different probability of being selected for mutation. The probability depends on the impact or feedback from the WUT. The impact will also influence how narrow the mutation range is.}
    \label{fig:hypermutation}
\end{figure*}

\paragraph{Adaptive hypermutation (AH)}
Doing simple mutation or using the response dictionary can be tricky for web API fuzzing because a WUT may treat each request parameter differently. For example, in a WUT, an \textit{id} field can be checked multiple times because it may relate to another data table, so it must be correct. Still, a \textit{name} may be automatically used without performing a deep examination, so using very varied names will always result in the same successful response from the WUT. Zhang et al. \cite{zhang_adaptive_2022} explained this situation in detail in which some parameters do not impact WUT execution flow. Spending too much time mutating them will be useless because it will not improve the effectiveness of the fuzzer. 

Therefore, inspired by gene mutations, they proposed an adaptive hypermutation to select and mutate genes adaptively based on their feedback and mutation history. This strategy is called hypermutation because it applies high mutation rates, leading to the mutation of multiple genes on average. One gen corresponds to one HTTP request parameter with either an atomic type (e.g., integer or boolean) or artificial types consisting of other genes (e.g., Date object consisting of day, month, and year). The feedback can be how many fuzzing objectives (e.g., code coverage) are achieved or the branch distance score (recall Section \ref{branch-distance}). Based on the mutation history, this strategy will only select genes that can help achieve the fuzzing objectives. The selected genes will be mutated differently depending on the type. For example, the mutated integer value will be calculated using formula \textit{a \textpm 2\textsuperscript{i}} which \textit{a} is the old value and \textit{i} is the random value within an adaptive range that depends on the impact score. For more detail, see an illustration in Figure \ref{fig:hypermutation}. The figure explains that only chosen genes, instead of all, will be mutated in the mutation process.

\paragraph{Attention-based models (AM)}
Lyu et al. \cite{lyu_miner_nodate} employs a trending machine learning method, an attention-based model, to help produce plenty of test cases that have more impacts on the WUT. They utilise such a model to learn the implicit relationship between parameter name-value pairs. They designed the model consisting of a Gated Recurrent Unit (GRU) neural network \cite{cho-etal-2014-learning}, an attention layer \cite{luong2015effective}, and a linear layer to result in valid param-value pairs for each request name. During the fuzzer runs, the collection module collects requests with valid responses to be set up as the training data for the model, in which a single row of the training data is in the form of \textit{$<request_name><param_value_1>..<param_value_n>$}. The fuzzer retrains the model every two hours from scratch.

\paragraph{Many Independent Objective (MIO)}
Arcuri \cite{arcuri_test_2018} proposed a fuzzing algorithm called Many Independent Objective (MIO) to help the web API fuzzing framework cover many objectives with a limited search budget. This algorithm is the refined version of the Whole Test Suite (WTS) \cite{fraser_whole_2013}, his previous algorithm. The new algorithm evolves the initial random population by applying rank selection, crossover, and mutation processes. Whenever branch coverage metrics are set as targets, the evolution process will be directed to cover all respective branch statements. The program instrumentation enables the model to detect covered branch statements because it registers and puts a probe in each scanned statement. If a registered branch is reached and the input can satisfy the branch condition, the branch is covered and gets the maximum heuristic score (h-score 1) in the fitness function \cite{rojas_detailed_2017}. Otherwise, the score is below 1, depending on how far the input is to solving a branch constraint. Since it is limited by time budget and has to cover many targets, the model will only focus on particular targets that can be solved in final iterations.

\subsubsection{Problem: More invalid requests}
Generating error cases can be tricky because the fuzzer cannot simply use random data. Since the malformed input is expected to test exceptional application scenarios, utilising the random data generator is insufficient to assess all possible error schemes. Fuzzer should introduce the faulty inputs slowly to reach deeper bug codes hidden in deeply nested conditionals. Most web API fuzzing frameworks adopt this idea and propose several strategies as follows.

\paragraph{Constraint violation (CV)}
Constraint violation techniques are the most commonly used techniques to create incorrect input and potentially cause WUT to crash. Fuzzing generally knows the constraints in filling valid values to the HTTP request, so at this time, it just ignores the constraint rules. This technique involves: 1) changing the data type of request parameters (e.g., from String to an array) \cite{van_rooij_webfuzz_2021}, 2) missing the required request parameters \cite{viglianisi_resttestgen_2020}, 3) violating the specific field's constraint (e.g., giving a long string for a field setting up the \textit{maxLength} constraint). However, implementing this strategy too aggressively in a single HTTP request can lead to "400 bad requests" because the server considers it a malformed HTTP request.

\paragraph{Rule-based schema mutator (RSM)}
Godefroid et al. \cite{godefroid_intelligent_2020} formulated some schema fuzzing rules for the schema mutator. The schema in this context is the HTTP request body consisting of parameter name-value pairs. This schema can be seen as a tree structure because certain parameters may have child parameters. The schema fuzzing rules involve the node rule, which defines the node modification techniques, and the tree rule, which determines which part of the schema will be fuzzed. The node rules cover dropping one child node, removing all nodes and keeping only one node, duplicating a new child node, and changing the node type. Then, the tree rules encompass picking a single node, choosing a path consisting of several nodes, or selecting all nodes, to apply to the node rules.

\paragraph{Vulnerability dictionary (VD)}
To trigger particular vulnerabilities, Gauthier et al. \cite{gauthier_experience_2022} fill specific HTTP parameter values with the values from a pre-defined dictionary of vulnerable payloads. This dictionary contains a set of vulnerable payloads mapping to certain vulnerability types. For example, when fuzzing requests a SQL injection payload for the name field, the dictionary will deliver strings like \texttt{'  OR  '1'='1' --}. Van Rooij et al. \cite{van_rooij_webfuzz_2021} employ a real-life XSS payload dictionary to bring their fuzzing framework effective malicious data. Zhao et al. \cite{zhao_cefuzz_2022} determined special seeds containing commands or functions (e.g., command 'system' in Linux) for triggering a PHP Remote Command/Code Execution (RCE) vulnerability. Based on the mutation rule, their fuzzing framework will take a particular seed to be combined with the formula forming the test case. Zhang et al. \cite{zhang_zokfuzz_2022} use the same concept, but the initial seeds contain SQL injection payload.

\paragraph{Tracked fault generator (TFG)}
Laranjeiro et al. \cite{laranjeiro_black_2021} employ a fault generator to insert fault codes into valid API requests guided by a Fault Mapper. The mapper tracks the inserted faults and their injection location in API requests to avoid exploring already visited locations. The fault dictionary produces concrete faults containing 57 fault rules, such as replacing valid values with null, removing random elements in an array, and others. \\

\noindent\fbox{%
    \parbox{\dimexpr\linewidth-2\fboxsep-2\fboxrule}{\textbf{Answer to RQ6}: To expand the existing input, most web API fuzzers use randomly modified values and dictionaries (either response or vulnerability dictionary). The others use corpus mutation, adaptive hypermutation, attention-based model, MIO algorithm, constraint violation, rule-base schema mutator, and tracked fault generator.}%
    % {\textbf{Answer to RQ6}: To expand the existing input, the most prominent solutions are using randomly modified values and dictionaries (either response or vulnerability dictionary). The others are corpus mutation, adaptive hypermutation, attention-based model, MIO algorithm, constraint violation, rule-base schema mutator, and tracked fault generator.}%
}%

\subsection{Choosing WUT for Experimental Evaluation}
\label{evaluation}
\noindent\fbox{%
    \parbox{\dimexpr\linewidth-2\fboxsep-2\fboxrule}{\textbf{RQ7}: What benchmarks are used for empirical evaluations?}%
}%
\\ \\
All fuzzing researchers conducted experimental studies using various benchmarks to evaluate whether the proposed works were better than the previous ones. Generally, the researchers test their fuzzers on public (online) web applications, local benchmarks they prepared, or third-party benchmarks.

\subsubsection{Public WUT}
\label{targeting-public-web}
Most prior studies (43\%) chose public web applications as the fuzzer target. Even though those applications can be relatively easy since a fuzzer user does not need to install and deploy them, they have some limitations: no source code and no instrumentation. No source code means the fuzzer cannot analyse the code workflow to generate better test cases, and no instrumentation means the fuzzer does not know what is happening during the execution of the WUT. Therefore, the WUT must be treated as a black box. Fuzzing researchers can consider public web applications listed in the \textit{APIS.Guru}\footnote{https://apis.guru/browse-apis/} website. Viglianisi et al. \cite{viglianisi_resttestgen_2020} used the website as the reference to test 87 web APIs, perform 2612 testing operations, and find 151 internal error faults. Furthermore, Laranjeiro et al. \cite{laranjeiro_black_2021}, who designed bBOXRT implemented in Java, tested 52 public REST services comprising 1,351 operations. The fuzzer detected at least one robustness problem (explained in Section \ref{specific-vulnerability}) in half of the services tested. 

On the other hand, some fuzzing papers considered popular and public RESTful Web API as their target. For example, Atlidakis et al. \cite{atlidakis_restler_2019} tested their fuzzer on some popular online RESTful APIs: three \textit{Azure}\footnote{https://azure.microsoft.com/en-us/} and one \textit{Microsoft Office365}\footnote{https://www.office.com/} online services. Their fuzzer found some bugs in those services that were 500-response codes. In addition, they also conducted an experiment on GitLab API\footnote{https://docs.gitlab.com/ee/api/} deployed on a local server and found bugs in commit and branch-related operation APIs. The other web API fuzzer, RESTest \cite{martin-lopez_restest_2021}, also performed similar experiments, evaluating its performance on some popular web applications: GitHub\footnote{https://github.com/}, Foursquare\footnote{https://foursquare.com/}, and others. It got bugs related to server and client error response codes. Another example is Peng et al. \cite{peng_automated_2022} tested their fuzzer to ByteDance\footnote{https://www.bytedance.com/en/}.

\subsubsection{Self-developed benchmarks}
\label{self-developed-benchmark}
Besides testing public services, the fuzzing researchers can evaluate their fuzzers on open-source web applications running on their local devices. Using such applications as the WUTs is more beneficial than testing the public closed-source services because fuzzing frameworks can access the source code. Analysing or instrumenting the code gives fuzzers much better knowledge, leading to higher code coverage and more bug-catching. Depending on the need, the researchers can utilise public artefact datasets from other researchers or build new ones. 
Therefore, developing a new benchmark containing some open-source web applications to test built fuzzers is a good option. Around 39\% of prior works did it. For example, EvoMaster Benchmark (EMB)\cite{arcuri_emb_2023} holding various web applications, such as Java, Kotlin, JavaScript, and C\# was developed to test EvoMaster \cite{arcuri_restful_2017}. On the other hand, the work of Van Rooij et al. \cite{van_rooij_webfuzz_2021} proposed an actual bug injection methodology into PHP-based web applications to create a proper benchmark. Getting inspiration from similar models like LAVA \cite{dolan-gavitt_lava_2016}, they gave a standard way to evaluate identical fuzzers in finding web vulnerabilities. Their benchmark contained CE-Phoenix\footnote{https://phoenixcart.org/}, Joomla\footnote{https://www.joomla.org/}, and others that had been injected with XSS vulnerabilities.

\subsubsection{Third-party benchmarks}
Since creating a new benchmark requires much engineering effort, the fuzzing researchers can use existing third-party benchmarks. Around 18\% of existing studies employ third-party benchmarks as their WUTs. The third-party benchmarks, also called test-bed applications, are designed by other people to contain many bugs for cyber security learning \cite{deepa_securing_2016}. Some examples of test-bed applications are \textit{WebGoat}\footnote{https://github.com/WebGoat/WebGoat} and \textit{Gruyere} \footnote{https://google-gruyere.appspot.com/}, which ones among the targets of KameleonFuzz \cite{duchene_kameleonfuzz_2014}. Those vulnerable web applications, built as web testing education, were proven to contain XSS bugs. Compared to other scanners, KameleonFuzz with LigRE \cite{duchene_ligre_2013} could detect more true XSS bugs in those applications. Another work doing similar things is BackREST \cite{gauthier_experience_2022}. It took test-bed applications built in Node.js platforms, such as \textit{Nodegoat}\footnote{https://github.com/OWASP/NodeGoat}. It also compared its performance to vulnerable scanners like Arachni\footnote{https://www.arachni-scanner.com/} and OWASP ZAP\footnote{https://www.zaproxy.org/} to detect SQL injection, command injection, XSS, and DoS vulnerabilities. The result was BackREST could catch the vulnerabilities more than the scanners. The last example is Witcher \cite{trickel_toss_2023}, which used known vulnerable applications as its fuzzer targets, consisting of 8 PHP-based web, 5 C-based firmware images, 1 Java-based web, 1 Python-based web, and 1 Node.js-based web application, with a total of 36 known vulnerabilities. The experiments showed that Witcher was better than Burp\footnote{https://portswigger.net/burp/vulnerability-scanner} and other tools in either bug finding or code coverage. Similarly, Cefuzz \cite{zhao_cefuzz_2022} and ZokFuzz \cite{zhang_zokfuzz_2022} used some PHP-based vulnerable web applications as their target, such as DVWA\footnote{https://github.com/digininja/DVWA}.\\

\noindent\fbox{%
    \parbox{\dimexpr\linewidth-2\fboxsep-2\fboxrule}{\textbf{Answer to RQ7}: For empirical evaluations, most studies (43\%) use publicly available WUT because the studies use black-box testing setup. Other works develop new benchmarks (39\%) or employ third-party benchmarks (18\%).}%
}%

\subsection{Summary}
Based on the explanations in prior sections, we summarise the information of the existing web API fuzzers in Table \ref{fuzzer-comparison-table2}. Instead of paper-based, we resume the insight based on the frameworks the researchers introduced in their works. We got 22 unique web API fuzzing frameworks even though we reviewed 53 papers because one framework can be explained or improved by multiple studies. For example, EvoMaster is used in nine studies \cite{arcuri_restful_2017}\cite{arcuri_evomaster_2018}\cite{arcuri_test_2018}\cite{arcuri_restful_2019}\cite{zhang_resource-based_2019}\cite{arcuri_enhancing_2022}\cite{zhang_adaptive_2022}\cite{golmohammadi_netc_2023}\cite{zhang_javascript_2023}. We sorted the frameworks based on the year they were published first. Most of them are black-box fuzzing, which only relies on HTTP responses. The others also utilise the HTTP response but are supplemented by other feedback, as listed in the table.

\begin{table*}
\caption{Web API fuzzer comparison.}
\label{fuzzer-comparison-table2}
    \begin{tabular*}{\textwidth}{@{\extracolsep{\fill}}p{3cm}|p{1cm}p{2cm}p{3.5cm}p{1.5cm}lp{2cm}@{}}
        \hline
         Fuzzer Name& Appr& WUT& Template \mbox{Generation}& Mutation& Feedback&  Vulnerability\\ \hline 
         KameleonFuzz \cite{duchene_kameleonfuzz_2014} & Vul \faIcon{square}& 3rd-party benchmark & HTML crawler \cite{duchene_ligre_2013} & VD & Fitness score & XSS \\
         EvoMaster \cite{arcuri_restful_2017} \faIcon {file-archive} & Std \faIcon[regular]{square}& JVM-based, NodeJS-based \cite{zhang_javascript_2023}, .NET-based \cite{golmohammadi_netc_2023}& \faIcon[regular]{book} Smart sampling \cite{arcuri_restful_2019} & AH+MIO & Branch distance & Crash\\
         Re-Checker \cite{woo_re-checker_2018} & Std \faIcon{square} & 3rd-party benchmark & Manual corpus & RMV & HTTP response & Unexpected behavior\\
         RESTler \cite{atlidakis_restler_2019} \faIcon {file-archive} & Std \faIcon{square}& Public web & \faIcon[regular]{book} Dependency graph & RSM \cite{godefroid_intelligent_2020} & HTTP response & Crash, Resource violation \cite{atlidakis_checking_2020}\\
         Jan et al. \cite{jan_automatic_2019} & Vul \faIcon{square}& 3rd-party benchmark& No dependency & RMV & XML message & XML injection\\
         RestTestGen \cite{viglianisi_resttestgen_2020} \faIcon {file-archive} & Std \faIcon{square}& Public web & \faIcon[regular]{book} Dependency graph & RD & HTTP response & Crash, Mass assignment \cite{corradini_automated_2023}\\
         bBOXRT \cite{laranjeiro_black_2021} \faIcon {file-archive} & Std \faIcon{square}& Public web & \faIcon[regular]{book} No dependency & VD & HTTP response & C.R.A.S.H.\\
         RESTest \cite{martin-lopez_restest_2021} \faIcon {file-archive} & Std \faIcon{square}& Public web& \faIcon[regular]{book} No dependency & DBPedia + IDL & HTTP response & Crash\\
         WebFuzz \cite{van_rooij_webfuzz_2021} \faIcon {file-archive} & Vul \faIcon{adjust}& PHP-based & HTML crawler & VD & Code coverage & XSS\\
         HsuanFuzz \cite{tsai_rest_2021} \faIcon {file-archive} & Std \faIcon{square}& 3rd-party benchmark & Pre-defined template & RMV & TCL & Crash\\
         UFuzzer \cite{huang_ufuzzer_2021} & Vul \faIcon[regular]{square}& PHP-based & No dependency & RMV & Taint feedback & Unrestricted file upload\\
         BackREST \cite{gauthier_experience_2022} & Vul \faIcon{adjust}& NodeJS-based & JS Crawler & VD &  Taint feedback & SQLi, CMDi, XSS, DoS\\
         RestCT \cite{wu_combinatorial_2022} \faIcon {file-archive} & Std \faIcon{square}& Public web& \faIcon[regular]{book}Tree-based structure & RD & HTTP response & Crash\\
         Cefuzz \cite{zhao_cefuzz_2022} & Vul \faIcon[regular]{square}& PHP-based & No dependency & VD & Taint feedback & PHP RCE\\
         Zokfuzz \cite{zhang_zokfuzz_2022} & Vul \faIcon{square}& PHP-based & No dependency & VD & HTTP response & SQLi, XSS\\
         MACROHIVE \cite{giamattei_automated_2022} \faIcon {file-archive}& Std \faIcon{adjust}& Microservice benchmark& Dependency graph \cite{giamattei_automated_2024} & RD & HTTP response & Crash\\ 
         WAPT \cite{auricchio_automated_2022} & Vul \faIcon{square} & 3rd-party benchmark & HTML crawler & VD & HTTP response & XSS, etc\\
         foREST \cite{lin_forest_2023} & Std \faIcon{square}& Public web& \faIcon[regular]{book}Tree-based structure & RD & HTTP response & Crash\\
         Witcher \cite{trickel_toss_2023} \faIcon {file-archive} & Vul \faIcon{adjust}& PHP-based, Python-based, Java-based, NodeJS-based, Ruby-based & HTML \& JS crawler & CM & Code coverage & SQLi, CMDi\\
         MINER \cite{lyu_miner_nodate} \faIcon {file-archive} & Std \faIcon{square}& Public web& \faIcon[regular]{book} Length-oriented selection & AM & HTTP response & Crash\\
         NAUTILUS \cite{deng_nautilus_nodate} & Std \faIcon{square}& 3rd-party benchmark & \faIcon[regular]{book} OpenAPI annotation & VD & HTTP response & SQLi, Privilege Escalation, etc\\
         Leif \cite{lei_bootstrapping_2023} & Std \faIcon{square}& Public web& FET inference& RMV & HTTP response & Crash \\
         \hline
    \end{tabular*}
    \begin{center}
        \item \begin{center} Appr = Approach, Std = Standard fuzzing, Vul = Vulnerability-driven fuzzing \end{center}
        \item \begin{center} \faIcon {file-archive} = open-source project, \faIcon{square} = black-box, \faIcon{adjust} = grey-box, \faIcon[regular]{square} = white-box, \faIcon[regular]{book} = utilize Open-API Spec\end{center}
        \item \begin{center} RMV = Randomly modified value, RD = Response dictionary, CM = Corpus mutation, AH = Adaptive hypermutation,\end{center}
        \item \begin{center}AM = Attention-based model, MIO = Many independent objective, CV = Constraint violation, VD = Vulnerability dictionary, RSM = Rule-based schema mutator, TFG = Tracked fault generator\end{center}
    \end{center}
\end{table*}

\section{Open Challenges}
\label{web-fuzzing-challenges}
\noindent\fbox{%
    \parbox{\dimexpr\linewidth-2\fboxsep-2\fboxrule}{\textbf{RQ8}: What open challenges are identified?}%
}%
\\ \\
Section \ref{review-result-restful-api} has explained a variety of prior approaches and techniques to enable fuzzing strategies for web application security testing. Despite the progress made, some shortcomings and challenges still need to be addressed. This section discusses open challenges in web API fuzzers mentioned in the summarized studies. The challenges are extracted from the relevant sections of each reviewed paper. The challenges mentioned in those papers but are tackled by more recent studies are excluded from the discussion.

\subsection{\{In\}Effectiveness of Instrumentation}
In many contexts, the use of grey or white-box web API fuzzers might provide better results for fuzzing effectiveness, assuming that access to the WUT source code is possible. However, using these fuzzing approaches may introduce an additional overhead due to the required preparation time (e.g., instrumentation) before testing. As explained in Section \ref{instrumentation-problem}, various methods of instrumentation have been developed over the years. However, whether the instrumentation pays off is a valid question. 

Generally, the instrumentation makes the source code bigger and slows down the server execution. It is caused by the probe code placed everywhere \cite{van_rooij_webfuzz_2021} in the WUT. In addition, the instrumentation process forces the WUT to be re-compiled and rebuilt, which certainly takes a significant amount of time. According to our preliminary experiments, carrying out fuzz testing for a short duration (e.g., 1 hour) is often ineffective. This is because the fuzzing process is only carried out after the re-compile and re-build process has been completed, and in some cases, the compile and build processes may take longer than the fuzzing process itself. The problem becomes more acute when the web developers are expected to deliver the application as quickly as possible, which may render the instrumentation process unacceptable. Therefore, the existing instrumentation techniques can be less effective in real-world settings where WUT consists of many files and the development is fast-paced. 

Designing a custom interpreter to catch interesting execution paths, rather than instrumenting the source code, can be a possible workaround to this problem because the majority of web applications are interpreter-based. Trickel et al. \cite{trickel_toss_2023} have applied this idea to augment the PHP interpreter tool for coverage analysis. However, there is still no experimental study that compares fuzzing with instrumentation and fuzzing without instrumentation (e.g., interpreter augmentation) in terms of effectiveness.

\begin{figure*}
    \centering
    \includegraphics[width=0.8\linewidth]{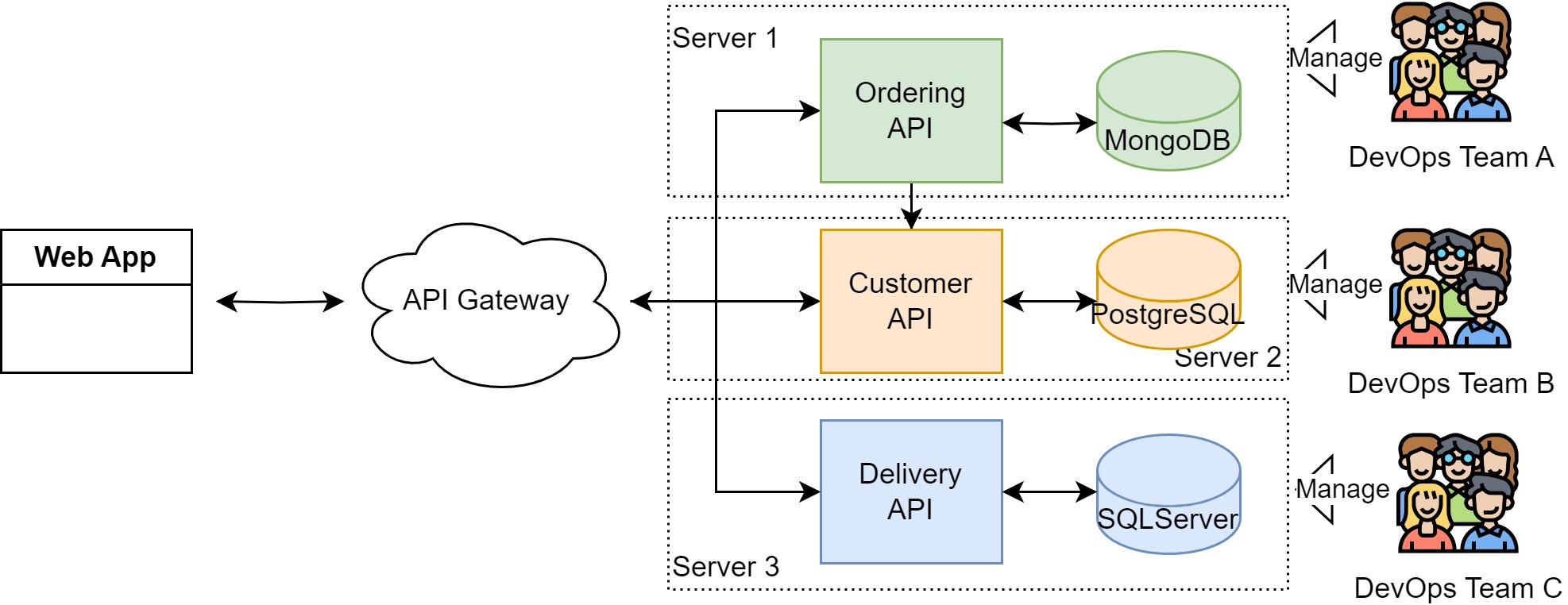}
    \caption{Illustration of a web application using the micro-service approach. Each service treats other services as black-box systems, which means it only sends inputs and receives outputs through provided communication channels and interfaces without having any internal information. Solely testing the API gateway can be misleading since the tester does not know which service crashes.}
    \label{fig:microservice}
\end{figure*}

\subsection{Complexity of Handling Microservice Architecture} %#3
Most world-leading web applications are using microservice architecture because it brings many benefits, such as ease of maintenance \cite{li_understanding_2021}. Since this kind of application is designed to be distributed, and each microservice typically operates independently, testing interactions between these distributed components can be complex \cite{blinowski_monolithic_2022} (see Figure \ref{fig:microservice}). This complexity presents unique testing challenges because the software tester must isolate each service properly to be able to distinguish the regions of the code where the problems arise. When only testing the main service without considering such complex interactions behind it, it may look as if the main service crashes often even though the real problem originates from other services that do not respond properly. In the microservice architecture, the main services exposing APIs to external users are called \textbf{edge microservices}; the others are called \textbf{internal microservices} because they expose their APIs only for internal services \cite{indrasiri2018microservices}. In addition, employing existing web API fuzzing techniques that are not designed for microservice applications may require the application testers to install and configure fuzzer for each service. It means that the fuzzer does not treat the WUT as a whole. The web API fuzzing framework should ideally be able to handle this enterprise architecture at scale.

Recently, researchers designed testing approaches to handle microservice architecture. For example, Giamattei et al. \cite{giamattei_automated_2022} proposed a grey-box strategy for automated testing and monitoring of internal microservices interactions. Their recent experiment \cite{giamattei_automated_2024} showed that their proposed framework delivered crucial information about internal coverage and failure, and inferred causality in failure chains. Zhang et al. \cite{zhang_white-box_2023} also developed a white-box fuzzer specially designed for RPC-based APIs that are commonly used in microservice architectures. However, both studies stated that further reearch needed to improve the fuzzers in this setting.

\subsection{The Difficulty of Testing Public WUT}
Most reviewed fuzzing papers use online web applications (e.g., Youtube API\footnote{https://developers.google.com/youtube/v3}) as the WUTs since they are popular and used by a large number of people. Even though preparing those WUTs for testing is easy, no installation or deployment is required prior to fuzzing, some challenges exist that make their testing difficult.

Firstly, since the majority of those popular web applications are closed-source, users can only test them using a black-box web API fuzzer, leading to code coverage ignorance. %\cite{martin-lopez_test_2019}. 
Without code coverage information, the fuzzers likely test only a small part of the WUT, which may render their results invalid. Secondly, popular web applications managed by big companies usually prohibit massive requests from the public \cite{wu_combinatorial_2022}. Besides determining low web resource quota, they also usually set up short-lived access tokens \cite{atlidakis_restler_2019} that require the users to do the authentication process more often. These two challenges make use of (online) closed-source WUTs for evaluating fuzzers less ideal.

Lastly, even if the researchers employ open-source WUTs (e.g., WordPress\footnote{https://wordpress.org/}), performing white-box fuzzing over these WUTs, that is, analyzing the source code to get more comprehensive understanding of the code, is difficult because those WUTs are often large projects with complex source codes \cite{trickel_toss_2023}. This situation requires the fuzzers to understand the recent programming features the WUT uses, such as object-oriented technique and MVC (model-view-controller) architecture. Therefore, more studies need to address this last challenge to improve existing web API fuzzers.

\subsection{Low-quality Corpus} %#1
Initial test cases play an essential role in all kinds of fuzzing since they are the initial corpus for the mutation process. Corpus in the web API context is the HTTP request sequence collection. If the corpus provides highly diverse, valid inputs, the work of the mutator can become easier because the mutator does not need to explore many more execution paths. As explained in Section \ref{review-result-restful-api}, existing web API fuzzing frameworks usually use only OpenAPI documents or HTML pages to create the initial corpus. Those documents are useful to help the fuzzer; however, some studies have shown that those documents may be incomplete. For example, the work of Deng et al. \cite{deng_nautilus_nodate} states that the information extracted from such documents is insufficient to produce diverse yet correct requests. Relying solely on these limited documents to generate an initial corpus for fuzzing can be risky because it produces low-quality corpus that will significantly affect the overall fuzzer performance.

Therefore, keeping the corpus small yet powerful is essential rather than  an extensive more complete corpus. The good corpus should be minimised before use, which means there should be a process to discard similar inputs that lead to same execution paths. The study from Herrera et al. \cite{herrera_seed_2021} shows that the minimised corpus can lead the fuzzer to explore new execution paths faster because it does not need to waste the  time executing inputs that produce an already known output. To our knowledge, most existing works on web API fuzzer do not perform the corpus minimisation process yet.

\subsection{Lack of Web API Fuzzing Benchmarks}
Researchers have developed various web API fuzzers using black-box, grey-box, or white-box approaches, and more are expected to come. These developers of fuzzing frameworks often claim that their fuzzers are really good at finding web vulnerabilities over a certain set of WUTs. However, there is neither a widely accepted benchmark for web API fuzzers nor established security-relevant metrics in comparing the performances of the fuzzers. For a proper and fair evaluation and comparison of fuzzers, comprehensive information about bug numbers detected by the fuzzers, including the miss and false bugs rate, is needed. Therefore, it is essential to develop a particular benchmark for web API fuzzers. Similar to binary fuzzing benchmarks (e.g., \cite{hazimeh_magma_2020}), the Web API fuzzing benchmarks should contain a diverse set of WUTs with injected bugs or vulnerabilities, integrate some of the well-known fuzzers and some scripts to automate all processes: deploying the WUTs, injecting the vulnerabilities, starting the fuzzers, and counting all evaluation metrics.

Even though there is no extensive web API fuzzing benchmark yet as in binary fuzzing, some benchmarks that have been made publicly available, as (explained in Section \ref{self-developed-benchmark}). For example, Arcuri et al. \cite{arcuri_emb_2023} released EvoMaster Benchmark (EMB), a set of comprehensive WUTs for evaluating EvoMaster fuzzer. However, the WUTs in EMB are just standard applications released by other developers, without annotated bugs inside them. Without the bug information, benchmark users cannot compare the web API fuzzer performance in catching bugs.
\\

\noindent\fbox{%
    \parbox{\dimexpr\linewidth-2\fboxsep-2\fboxrule}{\textbf{Answer to RQ8}: Identified open challenges are ineffectiveness of instrumentation, the complexity of handling microservice architecture, the difficulty of testing the public WUT, low-quality corpus, and lack of web API fuzzing benchmark.}%
}%

\section{Potential Research Directions}
Apart from the open challenges described in Section \ref{web-fuzzing-challenges}, we identified some potential research directions related to web API fuzzing. In doing so, we have taken into account the technologies that are currently being developed: web client programming, mobile web, and generative AI.

\subsection{Fuzzing for Web Client Programming}
Given the increasingly sophisticated web applications, the load on web servers has been growing. For example, recommendation services employing artificial intelligence methods in marketplace platforms are resource-hungry. The work of Wai et al. \cite{barolli_code_2022} investigates web client programming as one of the solutions to distribute the server load. This programming paradigm tries to move some of the computational processes from servers to client devices. This approach is reasonable since smartphones and computer desktops will constantly evolve and bring new hardware features to carry out sophisticated computations. Nowadays, while Java-script language can be considered as the main web client programming language, it is expected that more client programming languages that offer more portable and efficient features will be adopted in the future. For example, WebAssembly, developed by Haas et al. \cite{haas_bringing_2017} together with the W3C community to bring low-level C programming into the web, has attracted much attention from the fuzzing community. Therefore, employing fuzzing for this application type can become popular in the future.

\subsection{Fuzzing Mobile Web Applications}
Many web pages are prepared/delivered explicitly for/to mobile devices \cite{serrano_mobile_2013}. Even though mobile web applications are generally similar to regular web applications, there are certain differences. Both usually have the same content sources, but they are rendered differently depending on the client device's capabilities. For example, the mobile web does not show complex user interfaces in order to provide better user experiences on a smaller screen. Next, some mobile hardware limitations may cause the mobile device not to execute complex client-side (e.g., Javascript) code. Considering the growth of the  smartphone market, many more mobile web applications are expected to be launched. Testing this type of WUTs is challenging because fuzzers often demand more sophisticated computational resources that the mobile devices do not have. Developing a fuzzing approach for smartphones or emulating the mobile web on a desktop can be a possible alternative even though there are certain drawbacks of this approach. For example, mobile web emulation can be less accurate since mobile devices are highly diverse, from old to cutting-edge systems. Whatever approach is chosen, anticipating this technology as early as possible is a reasonable consideration.

\subsection{Generative AI for Web Testing}
Generative artificial intelligence (AI) is a class of AI techniques and models that can generate new content, such as text, images, music, or other forms of data \cite{noauthor_generative_nodate}. These models are designed to learn the patterns and structures present in the training data and then produce outputs with similar characteristics. Generative AI, a branch of machine learning, has become increasingly popular in diverse domains including software testing as it can generate content that was not explicitly present in the training data. 
% Considering its trend and features, employing this technique for software testing is an interesting research direction. 
Some reviewed works in web API fuzzing already utilised powerful AI techniques partially in some steps. For example, Lyu et al. \cite{lyu_miner_nodate} employed the attention model for filling values in the template rendering stage. It is estimated that more papers will adopt generative AI techniques fully to generate HTTP requests to test web API in the future.

\subsection{Support for Diverse Web Security Vulnerabilities}
In addition to using the existing OWASP vulnerability list \cite{noauthor_owasp_nodate} \cite{noauthor_owasp_nodate_api} to define vulnerabilities to be detected, developing strategies for the detection of new bug types that are different from the well-known ones is a promising research direction. Since the OWASP list is based on past investigations, imagining far ahead about potential faults that may not have happened yet is essential, especially by considering/predicting user habits and bugs that may occur in the future. For example, the work of Atlidakis et al. \cite{atlidakis_checking_2020} identified a new vulnerability family that can result with the hijacking of a WUT and developed a method to detect bugs related to this vulnerability. They defined four new security rules related to web resource management: use-after-free, resource-leak, resource-hierarchy, and user-namespace rules. If the web API violates one of those rules, the authors conclude that the WUT has this vulnerability. 

Extending the existing bug criteria can also be a good option besides specifying completely new security bugs. This is also important in enabling security practitioners to catch complex bugs easily. For instance, the work of Pan et al. \cite{pan_edefuzz_2024} developed a strategy to detect Excessive Data Exposure (EDE) vulnerability. Even though OWASP included this vulnerability, catching this bug is not easy since it does not trigger a crash or an unexpected behaviour. Therefore, determining more precise test oracles will help fuzzers to recognise the vulnerability.
Finally, among the vulnerabilities covered by the OWASP list, only two types, SQL/code injection and XSS, have been studied extensively. Effective detection strategies for the other vulnerabilities are still quite open to research.

\section{Threats to Validity}

\paragraph{Paper selection}
Choosing relevant papers from search databases and filtering irrelevant studies requires manual effort and expertise. As a result, this process might be susceptible to human error. The authors did searching and filtering of the paper collection more than two times to reduce that possibility.

\paragraph{Information extraction}
Analysing and extracting relevant information from the selected papers was done manually too. Although the authors are confident with the results, they might also be susceptible to human misunderstanding. To reduce this chance, the authors read and analysed the papers more than two times.

\section{Conclusion}
In this survey paper, we reviewed 53 articles presenting research results on web API fuzzing frameworks, and classified them according to their testing objectives and techniques used to generate HTTP requests, utilise feedback from the WUT, and mutate existing requests. In addition, some insights about open challenges and anticipated research related to web API fuzzing are provided. Ultimately, we aim this paper can help the readers choose which techniques to improve for developing a better web API fuzzing framework.

\section*{Acknowledgements}
\paragraph{Author Contributions}
The first author conducted the literature survey and wrote most of the paper; the second author contributed to the discussions and reviewed the paper; and the third author contributed to the discussions, writing and reviewing of the paper.

In addition, the authors thank \href{https://orcid.org/0000-0002-5331-4033}{Tariq Bontekoe} for proofreading this article.

\paragraph{Funding}
The first author has received scholarship funding from the Center for Financing of Higher Education (BPPT) and the Indonesia Endowment Fund for Education (LPDP) under the Indonesian scholarship schema.

\section*{Declarations}
\paragraph{Conflict of Interest}
The authors do not have any financial or non-financial interests to disclose that are relevant to the content.

\paragraph{Ethical Approval} Not applicable.

\printbibliography

\clearpage
\appendix

\section{OpenAPI Specification Example}
\label{appendix:openapi-example}
\begin{lstlisting}[language=Python, firstline=1, lastline=33, numbers=left]
openapi: "3.1.0"
info:
  title: API Example
paths:
  /book:
    get:
      parameters:
        - name: limit
          in: query
          description: How many items to return at one time (max 100)
          required: false
          schema:
            type: integer
            maximum: 100
            format: int32
      responses:
        '200':
          description: A paged array of books
          headers:
            x-next:
              description: A link to the next page of responses
              schema:
                type: string
          content:
            application/json:    
              schema:
                $ref: "#/components/schemas/books"
        default:
          description: unexpected error
          content:
            application/json:
              schema:
                $ref: "#/components/schemas/Error"
\end{lstlisting}

\newpage
\section{Grammar example generated by RESTler \cite{atlidakis_restler_2019}}
\label{appendix:grammar-example-by-restler}
\begin{lstlisting}[language=C, firstline=15, lastline=24, numbers=left]
Request(
    static("GET /api/customer/data"), 
    static("?idcustomer="), 
    fuzzable("integer"), 
    static("&name="), 
    fuzzable("string"), 
    static("HTTP/1.1"), 
    static("Accept:application/json"), 
    ...... 
)
\end{lstlisting}

\end{document}